# FeatureSense: Protecting Speaker Attributes in Always-On Audio Sensing Systems

BHAWANA CHHAGLANI, University of Massachusetts Amherst, USA
SARMISTHA SARNA GOMASTA, University of Massachusetts Amherst, USA
YUVRAJ AGARWAL, Carnegie Mellon University, USA
JEREMY GUMMESON, University of Massachusetts Amherst, USA
PRASHANT SHENOY, University of Massachusetts Amherst, USA

Audio is a rich sensing modality that is useful for a variety of human activity recognition tasks. However, the ubiquitous nature of smartphones, wearables, and smart speakers with always-on microphones has led to numerous privacy concerns and a lack of trust in deploying these audio-based sensing systems. This paper addresses this critical challenge of preserving user privacy when using audio for sensing applications while also maintaining utility. While prior work focuses primarily on protecting recoverable speech content, we show that sensitive speaker-specific attributes such as age, gender, and ethnicity can still be inferred after masking speech and propose a comprehensive privacy evaluation framework to characterize this speaker information leakage. Next, we design and implement *FeatureSense*, a lightweight open-source library that provides a set of generalizable privacy-aware audio features that can be used for a wide range of sensing applications. Additionally, we present an adaptive task-aware feature selection algorithm that optimizes the privacy-utility-cost trade-off based on the application requirements or context. Through our extensive evaluation, we demonstrate the high utility of *FeatureSense* across a diverse set of sensing tasks, including environment sound classification, cough detection, and urban sounds detection. Our system outperforms existing privacy techniques by 60.6% in preserving user specific privacy. This work provides a foundational framework for ensuring trust in audio sensing applications by enabling effective privacy-aware audio classification systems, while minimizing the feature extraction cost.

## 1 INTRODUCTION

Audio sensing has become integral in supporting numerous applications, ranging from smart home devices and virtual assistants to health monitoring systems and security solutions[1–5]. Audio sensing can enable activity

Authors' addresses: Bhawana Chhaglani, University of Massachusetts Amherst, USA, bchhaglani@cs.umass.edu; Sarmistha Sarna Gomasta, University of Massachusetts Amherst, USA, sgomasta@umass.edu; Yuvraj Agarwal, Carnegie Mellon University, USA, yuvraja@andrew.cmu.edu; Jeremy Gummeson, University of Massachusetts Amherst, USA, gummeson@cs.umass.edu; Prashant Shenoy, University of Massachusetts Amherst, USA, shenoy@cs.umass.edu.

detection [3, 6–8], emotion recognition [9], gesture detection, health monitoring [1], and environment monitoring [10] given that many of these activities have discernible audio signatures that ML models can be trained on. However, audio-based sensing also raises significant privacy concerns [11]. Audio is a rich modality that can not only leak sensitive information such as speech content [11], but also a person's biometric identity, age, gender, ethnicity, health or emotional status [9, 12]. For example, there have been prior reports where accidental triggering of voice assistants has recorded private conversations [13, 14]. Given the proliferation of always-on microphone-equipped devices such as smartphones, smart speakers, and baby monitors [15], privacy concerns resulting from the use of audio for a range of sensing tasks need to be addressed.

Recognizing these privacy challenges with audio-based sensors, numerous approaches have been proposed to address them. Existing approaches include sub-sampling [16], filtering [10], shredding [17], speech obfuscation [18], eliminating likely speech segments [1, 11], and differential privacy [19]. In general, these techniques focus on suppressing human speech to ensure the privacy of the sensing task. For example, Pdvocal [20] uses non-speech body sounds for privacy-preserving Parkinson's detection. Crowdotic [19] uses non-speech audio for privacy-aware occupancy prediction. However, human speech is not the only private information present in an audio signal—raw audio can also reveal information about the speaker's identity and whereabouts. Voice data can disclose sensitive demographic details such as gender, age, ethnicity, health conditions, or emotional state, leading to unwanted profiling. This leaked data can reveal patterns like stress levels, accents, or socioeconomic status, which could be used without consent for targeted advertising or discriminatory practices. For example, insurance companies and health analytics platforms can use inferred demographics alongside cough frequency or acoustic biomarkers to categorize users into risk groups, potentially influencing policy offerings or premiums. In smart home settings, identifying whether a child or elderly person is home alone can make households vulnerable to targeted intrusions or manipulative marketing. Moreover, advertising systems have been shown to target users differently based on perceived gender or ethnicity, raising concerns of discrimination and bias. These demographic inference attacks are often easier to construct compared to full speaker identification attacks, which typically require large databases of speaker signatures. Even without explicit identity recovery, demographic inference can significantly narrow down or disambiguate individual speakers within multi-occupant or public environments, leading to privacy breaches through partial profiling. Thus, similar to speech leakage, leakage of speaker-related information can lead to personal harm, loss of trust, and broader societal risks.

To address these issues, our work presents a privacy-aware audio sensing approach that is based on a more holistic definition of audio privacy that emphasizes speaker leakage in addition to other risks. Speaker leakage often arises from the presence of speaker-specific acoustic cues in an audio signal that are unintentionally revealed. Even non-speech audio signals can reveal significant information about an individual. For example, Zhao et al. [21] has shown the feasibility of using breath sounds as a biometric for speaker recognition. Characteristics such as vocal tract resonance, pitch, speaking style, and breathing sounds can lead to identifiable information. Studies have shown that these features, even when captured in the background of recordings, can be used to infer speaker identity or characteristics like gender, age, and emotional state [22]. In our work, we measure and prevent speaker-related information leakage to provide more comprehensive privacy protection. Similarly to the word error rate (WER) that can be used to assess speech leakage, we propose a metric, the speaker information leakage index (SILI), which helps evade speaker leakage of a given system by considering important speaker demographic parameters using the CommonVoice dataset [23]. To prevent speaker leakage, we focus on the use of derived audio features, instead of raw audio, for various audio sensing tasks. Our hypothesis is that using carefully curated audio features, we can effectively perform a wide range of non-speech speaker-invariant audio classification tasks (like environment sound classification) while preserving speaker identity. Our approach is based on the observation that most *non-speech* audio sensing tasks, such as sound classification, anomaly detection, and event recognition, can be effectively performed using extracted audio features instead of the complete audio waveform. By limiting access to raw audio and focusing on task-relevant features, privacy risks can be significantly reduced.

However, identifying features that maximize task performance while minimizing sensitive information leakage remains a critical challenge. There is a pressing need for robust approaches that can effectively balance utility and privacy in audio sensing applications. This creates an opportunity to design systems that extract only the necessary information for the application while discarding or obfuscating irrelevant and sensitive content.

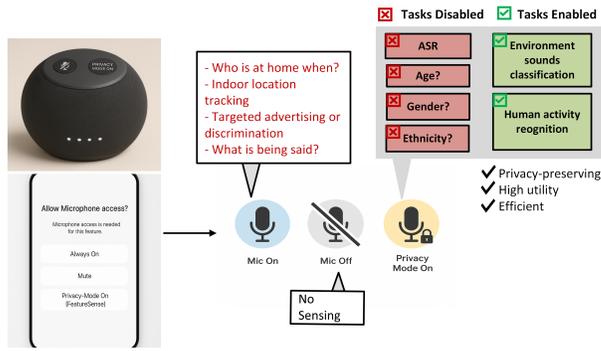

Fig. 1. Introducing privacy-mode on feature on microphone-equipped devices

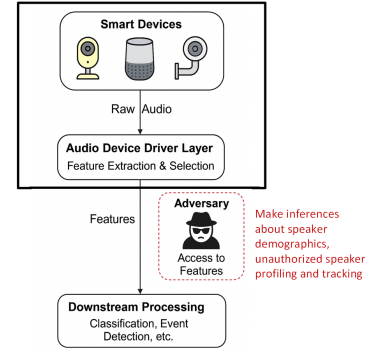

Fig. 2. *FeatureSense* attack model

Although there are some existing efforts [1, 24] that use privacy-conscious audio features for various applications such as cough detection, they require hand-crafted audio features. Our work seeks to eliminate this feature engineering effort which requires domain knowledge by automating feature selection for various tasks. To do so, we propose a privacy-aware audio feature library that can be used for a broad range of non-speech sensing tasks, while preserving user privacy and maintaining the effectiveness of audio classification. Our approach seeks to expose a set of features, instead of raw audio, using our proposed *FeatureSense* library, which consists of time domain, spectral domain, time-frequency, perceptual, voice-specific, statistical and derived features. *FeatureSense* is designed using carefully curated features that do not leak privacy, drawing from the source-filter model of speech production [25] and information theory. We show that our proposed features are well suited for a range of non-speech speaker-invariant sensing tasks. A key challenge however with using featurization-based approaches is the added computational cost of extracting these features in real time, leading to a *privacy-utility-cost* tradeoff. To further address this challenge and enable context-aware applications, we present an adaptive task-specific feature selection algorithm that optimizes the subset of features for the given application and its requirements. Our algorithm provides parameters that can be tuned based on the application constraints, resulting in an adaptive feature list. We show the effectiveness of our algorithm using various case studies: smart speaker-based environmental sensing in privacy mode and privacy-aware cough detection in health apps. We perform extensive evaluation to show that our approach preserves user privacy, while resulting in high utility and low cost. We compare the proposed system with existing privacy techniques and achieve 60.6% reduction in speaker leakage. We demonstrate high accuracy of 81.2% on ESC-50 [26], 85.3% on AudioSet [27], and 97.1 % UrbanSounds8k [28], thus showcasing high utility across various tasks. *FeatureSense* requires only 10ms to process features for 500ms of audio, making it well-suited for real-time applications with minimal computational overhead. Through case studies, we show that our adaptive feature selection algorithm can enable developers to design their application without the manual effort of feature engineering or requiring domain knowledge. In summary, we make the following contributions:

- **Comprehensive Speaker Leakage Evaluation:** We propose a novel metric SILI to evaluate speaker attribute leakage, addressing privacy gaps in existing metrics like WER which focus on speech recognition only. This helps in quantifying privacy leakage from feature-based representations. We conduct a systematic, comprehensive assessment of speaker demographic leakage (age, gender, ethnicity) using the proposed SILI, providing deeper insights into feature-level privacy vulnerabilities. Through extensive evaluation, we demonstrate the effectiveness of our approach in preventing speaker leakage as compared to the state-of-the-art techniques (including PrivacyMic [29], Samosa [16], Synthetic Sensors [30], and Kirigami [11]).

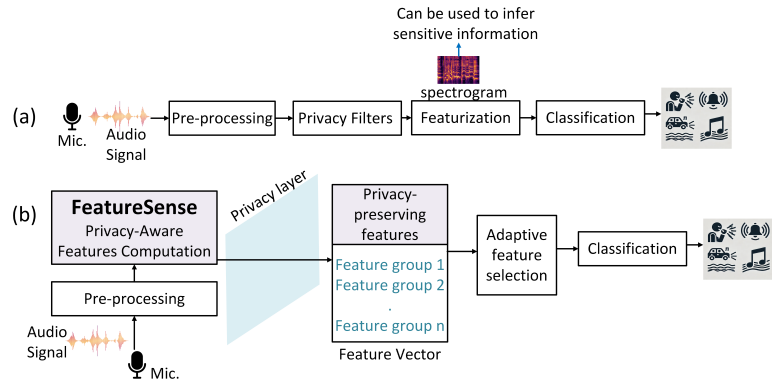

Fig. 3. (a) Existing Privacy Approaches and (b) Our Proposed Approach

- **Privacy-Utility-Aware Feature Library**: We design a general-purpose open-source library of privacy-aware audio features. We systematically identify a subset of low-level audio features that eliminate essential speaker- and speech-specific leakage while maintaining high utility for ambient event classification and human activity recognition.
- **Automatic Task-Aware Feature Selection:** We present adaptive task-specific optimization algorithm for automated feature selection that balances privacy, utility, and computational cost, based on task requirements and device constraints. This allows non-expert developers to deploy efficient privacy-aware sensing without requiring deep acoustic expertise.
- **Extensive Evaluation in Real-World Case Studies:** We evaluate FeatureSense on UrbanSound8k and AudioSet through two privacy-critical applications—smart speaker environmental sensing and cough detection in health apps—highlighting significant reductions in age and ethnicity leakage while maintaining high utility.
- **Lightweight and Deployable Privacy Layer:** Unlike prior approaches requiring complex models (e.g., adversarial training, differential privacy) or hardware changes, FeatureSense operates at the feature extraction layer and can be integrated into the device driver stack. We demonstrate sub-15 ms end-to-end latency on 500ms audio windows, making the system practical for always-on sensing on edge devices.

By addressing the limitations of current approaches and providing a more robust solution, *FeatureSense* represents a significant step forward in the field of privacy-preserving audio sensing.

## 2 BACKGROUND

### 2.1 Privacy Implications of Audio Sensing

In recent years, audio sensing technologies have become an integral part of our daily lives. However, the proliferation of these audio sensing technologies has raised significant privacy concerns. The always-on nature of many devices means that users are potentially being recorded at all times, even when they are unaware. This continuous monitoring can inadvertently capture sensitive personal information, including private conversations, health-related sounds, or background noises that reveal location or activities [22]. Advanced audio processing techniques can identify individuals based on their voice characteristics, potentially enabling unauthorized tracking or profiling. Moreover, voice analysis can reveal emotional states [9], stress levels, and even health conditions, which users may not wish to disclose. The risk of data breaches adds another layer of concern, as compromised audio data could lead to significant privacy violations and potential misuse of personal information. Thus, there is a need for prevent leakage of sensitive information from audio data to protect privacy.

## 2.2 Privacy-Preserving Techniques For Audio

Current approaches to audio privacy often rely on subsampling, frequency-based filtering, adding noise, removing/replacing speech segments, and differential privacy. Kumar et al. [17] propose sound shredding (randomly shuffled sound frames) and sound sub-sampling for privacy-preservation. This work shows that the shuffled sound recording makes it difficult to recover the text content of the original sound recording, yet we show that some acoustic features are preserved which retains the accuracy of context recognition. SaMoSa [16] uses sub-sampled audio (rates <= 1 kHz) to reduce intelligibility of the spoken content. PrivacyMic [29] removes all the audible frequencies from 300-8000 Hz, while FlowSense [10] uses optimal low-pass filtering to ensure privacy preservation. Liu et al. [31] examines the degree to which low-frequency speech ensures verbal privacy. This work showed that while low-frequency recording shows promise in preserving privacy by obstructing intelligible speech, it is not a comprehensive solution. SILENCE [32] selectively obscures short-term details, without damaging the long-term dependent speech understanding performance. A common approach is to identify speech segments in the incoming audio signal and then eliminate [1, 11] or obfuscate them [18, 33]. Liaqat et al. [18] distort speech to a greater degree than other sounds by identifying voiced segments. Similarly, Chen et al. [33] identifies vocalic regions and replacing the vocal tract transfer function of these regions with the transfer function from pre-recorded vowels. Recent work Kirigami [11] uses a lightweight, machine-learning-based speech filter that detects and removes human speech from audio data to prevent speech leakage. The effectiveness of these techniques rely heavily upon the speech/non-speech detection model. Additionally, current approaches consider human speech as the key private-sensitive information that needs to be protected, while our work considers privacy holistically by considering privacy leakage that goes beyond speech leakage. This work looks at privacy beyond speech by focusing on speaker leakage and presents a new audio sensing framework for protecting speaker privacy.

## 2.3 Common Audio Feature Representations

Conventional audio classification systems typically involve two main stages: feature extraction and classification as shown in Figure 3. Features like Mel-frequency cepstral coefficients (MFCCs) are extracted from audio signals, which are then input into machine learning models for classification into predefined categories. Existing efforts mainly use higher-dimensional representations derived from transformations of the audio signal like MFCC, Mel Spectrogram, Short-time Fourier transform (STFT) [4], or other feature representations and feed them to a deep learning (DL) model for classification. However, others have shown that these features do not maintain privacy, as they not only reveal speech, but also inflection, and prosody [34]. Further, these features can leak other essential speaker information due to their higher information density and richer representation of the signal. In contrast, our approach focuses on task-relevant, low-dimensional features (e.g., ZCR, RMS, spectral flatness) that inherently abstract the signal. This approach reduces computational overhead, limits privacy risks, and maintains effectiveness for specific applications, unlike high-dimensional features, which retain richer representations but pose greater privacy challenges.

**Application-specific Handcrafted Features**: Wyatt et al. have devised audio features that can successfully hide speech intelligibility, while simultaneously providing cues for prosody and recognition of conversations [34]. Other prior works have also designed task relevant features for a specific application to ensure privacy. For example, AeroSense [3] uses task-specific audio features to get privacy-preserving classification for the desired task. Similarly, Parthasarathi et al. [35] investigated a set of privacy-sensitive features for speaker change detection. Symdetector [36] proposed audio features to distinguish sound-related respiratory symptoms. Larson et al. [37] uses cough specific features to accurately distinguish cough from similar-sounding sounds. These techniques involve manual feature engineering to optimize for the accuracy of the task. We hypothesize that by using a set of generalizable privacy-aware features, we can effectively classify variety of audio classes.

## 2.4 Speaker Leakage

Speaker leakage is an important and largely unaddressed challenge in the domain of audio signal processing and privacy preservation. By speaker leakage, we mean the leakage of speaker demographic information such as their age, gender and ethnicity from an audio signal [38]. Studies have explored the use of breathing and non-verbal sounds for speaker identification [21]. Shen et al. [39] shows that gender can be inferred from non-linguistic audio data. Efforts like the Voice Privacy Challenge (2020) [40] aim to develop methods that balance speaker anonymization with utility for downstream tasks. Voice anonymization has been mainly used for speaker-independent speech recognition. However, voice anonymization techniques must effectively hide speaker-specific characteristics without degrading the audio quality or impairing its usability for downstream tasks like speech recognition. Additionally, many real-world systems require real-time processing, making it challenging to implement computationally intensive anonymization algorithms.

**Assessing Speaker Leakage**: The lack of universally accepted metrics and benchmarks for evaluating anonymization effectiveness makes it difficult to compare and prevent speaker leakage. The current method to evaluate the effectiveness of an audio privacy technique is subjective human intelligibility [10, 31] or automatic speech recognition (ASR) performance [11]. Although these techniques are useful for measuring speech leakage and quantifying it, the remaining signal could still have sensitive speaker-related information. Speaker identification systems are typically evaluated using metrics such as Equal Error Rate (EER), which balances false acceptance (FAR) and false rejection (FRR). Additional metrics like Accuracy, Detection Cost Function (DCF), and F1-score are used in forensic and authentication scenarios. However, while these assess speaker distinction, no standard metrics exist for evaluating speaker attribute leakage—unintended disclosure of characteristics like gender, age, or emotional state—posing challenges in quantifying and mitigating privacy risks.

## 3 MOTIVATION: WHY PREVENTING SPEAKER DEMOGRAPHIC LEAKAGE MATTERS

While inferring speaker demographic attributes such as age, gender, and ethnicity may seem benign, they can be misused in subtle yet harmful ways. Passive audio recordings—even without speech content—can enable profiling, targeted advertising, or biased treatment in both commercial and healthcare settings. Smart devices that suppress speech may still leak demographic traits via low-level features. Table 1 presents real-world scenarios where such leakage can lead to unintended privacy violations, highlighting the need for sensing systems that also guard against speaker attribute inference. FeatureSense targets audio sensing applications that rely on acoustic patterns but do not require speaker identity or speech content—such as ambient sound detection, pet activity trackers, human activity recognition, etc. In these contexts, detecting events like siren or coughing is essential, but inferring speaker age, gender, or identity is unnecessary and invasive. FeatureSense enables on-device sensing that preserves utility while minimizing speaker-related privacy risks.

FeatureSense: Protecting Speaker Attributes in Always-On Audio Sensing Systems • 7

### 3.1 Example Attack Scenarios

*Case 1: Health Apps Profiling Users from Cough Sounds*: Smartphone apps like Hyfe AI [41] detect coughs to monitor health, but often process raw audio that reveals speaker demographics. Such inferences could lead to targeted advertising, health-based discrimination, or insurance premium adjustments based on perceived risk profiles. FeatureSense eliminates this risk by enabling cough detection using only non-identifying acoustic features.

*Case 2: Smart Speakers Revealing Who's Home*: Devices like Alexa Guard and Google Nest monitor for emergency sounds (e.g., glass break, alarms) but also capture ambient audio that may reveal whether a home is empty, whether a child or elderly person is alone, or if someone is distressed. Such information could be misused for burglary targeting, surveillance, or profiling vulnerable individuals. FeatureSense enables detection of critical events without exposing speaker traits. The manufacturers of these devices can add FeatureSense to gain more trust with the users since the parent companies have advertising based business models.

*Case 3: Covert Tracking via Indoor Microphone Sensors*: Sensors with microphones are being deployed in hotels (e.g., voice assistants) and buildings to (e.g., Smart Thermostat [42]) enable smart building applications. However, they can also be susceptible to demographic leakage across space and time. When multiple speaker attributes are leaked and aggregated across different sensors and time points, adversaries could theoretically disambiguate individual speakers and reconstruct their movement patterns within a building, even without explicit speech recognition. Such covert surveillance raises serious concerns around privacy violations, discriminatory profiling, and unauthorized tracking. FeatureSense restricts data collection to privacy-safe features, enabling occupancy insights without identity exposure.

### 3.2 Real-World Incidents of Audio-based Speech and Speaker Leakage

Leakage of speech content and speaker attributes such as age, gender, or identity has led to real-world privacy breaches and legal consequences. In one incident, a user was inadvertently sent over 1,700 audio recordings from another household, exposing private conversations and identities [43]. Regulators have also taken legal action against major tech platforms for retaining children's voice recordings and location data without proper consent, violating child privacy protections [44]. Additionally, class-action lawsuits have revealed that some voice assistants recorded sensitive and confidential conversations without user knowledge or permission, leading to multi-million dollar settlements [45]. Such cases underscore the urgent need for privacy-preserving mechanisms like *FeatureSense* that proactively eliminate demographic and speech leakage at the feature level, ensuring responsible deployment of audio sensing technologies.

Table 1. Example Applications, Privacy Risks, and Their Implications

| Scenario (Device) | Potential Adversary | Privacy Risk (Potential Inferences) | Implication |
| --- | --- | --- | --- |
| Cough Monitoring Apps (e.g., Hyfe AI) | App vendors, insurance companies | (What is the speaker's age, gender and ethnicity?) (Can the speaker's identity be inferred?) | Targeted advertising, health-based discrimination, insurance premium manipulation |
| Smart Speaker Event Detection (e.g., Alexa Guard, Google Nest) | Cloud services, smart home platforms, rogue employers of companies | (Is a child left home alone?) (Is an elderly person present?) (Is the home empty?) | Security risks, burglary targeting, exploitation of vulnerable household members |
| Indoor Building Monitoring (e.g., Airport, Office & Hotel Mic Arrays) | Surveillance vendors, facility managers | (Did employee arrive late to work today?) (Are hotel occupants a couple in a secret relationship?) | Covert surveillance, group profiling, privacy violations in public/private spaces |

These examples highlight the need for privacy-aware feature selection mechanisms such as FeatureSense, which can mitigate sensitive inferences at the source without compromising application utility.

### 3.3 Attack Overview and Threat Model

**Privacy Leakage:** FeatureSense targets leakage of speaker-specific information in ubiquitous audio sensing systems. Even in applications that do not require speech recognition—such as cough detection, glass break sensing, or ambient sound monitoring—low-level acoustic features can inadvertently encode demographic traits (e.g., age, gender, ethnicity) and speaker identity. These leaks may occur despite the absence of explicit speech content. If exposed, they can be used for profiling, surveillance, or targeted decision-making. Our goal is to ensure that such speaker-related attributes remain private and cannot be inferred from the shared features.

**Adversary's Capability:** As shown in Figure 2, we consider a passive adversary who resides on the cloud server, third-party analytics platform, or within the model inference pipeline. The adversary does not have access to raw audio or metadata but only sees the extracted feature representations. Leveraging pretrained models, external datasets, and auxiliary knowledge, the adversary attempts to infer sensitive speaker traits from these features. We assume that feature extraction and selection occur securely and locally on the device (at OS-level), and that communication to the server is encrypted. The adversary does not modify or interfere with the sensing pipeline.

**Attack Scenarios**: We envision two primary attack scenarios: (1) *Demographic Inference:* The adversary attempts to infer the speaker's age, gender, or ethnicity from non-verbal audio features to enable profiling or targeted advertising. (2) *Identity Reconstruction:* Across sessions or locations, the adversary aggregates feature embeddings to re-identify users or track individuals over time.

**Defense:** FeatureSense mitigates these risks by identifying and eliminating features that leak speaker or speech-specific information using statistical, ML-based, and information-theoretic methods. It provides a tunable privacy-utility-cost framework and introduces SILI/CSLI metrics to evaluate and minimize leakage, enabling secure and responsible deployment of audio sensing applications. In our envisioned deployment, feature extraction and selection are performed locally at the device driver layer, prior to any transmission or cloud-based processing. This ensures that only compact, privacy-preserving audio features are shared with downstream services or remote servers.

### 3.4 *FeatureSense* Intended Use-cases

Many existing ubiquitous sensing platforms, including cough monitoring apps, smart speaker systems (e.g., Amazon Alexa Guard, Google Nest Aware), and automotive driver monitoring technologies (e.g., Tesla, Comma.ai) rely on continuous ambient audio collection. However, these applications primarily require detection of non-verbal events such as coughs, yawns, or glass breaking and do not need access to raw speech content or speaker demographic information. Without proper safeguards, adversaries can infer sensitive attributes such as user age, gender, or ethnicity. By integrating privacy-aware feature selection like FeatureSense at the device driver layer, these systems could continue to function accurately while guaranteeing protection against speaker and demographic leakage. This enables "always-on" sensing with explicit privacy guarantees, enhancing user trust and enabling new modes of operation (e.g., "privacy-preserving monitoring").

## 4 PRIVACY VULNERABILITIES IN EXISTING PRIVACY TECHNIQUES

Privacy vulnerabilities in audio sensing arise from the sensitive nature of audio data, which often contains identifiable information or unintended signals. Privacy in audio has many facets across both content and context [46]. As privacy varies with context, evaluating privacy leakage is non-trivial. Audio contains more privacy invasive information than just speech. So, there is a need to look at a more holistic definition of privacy, thus requiring expansion of privacy evaluation mechanisms. To comprehensively assess the privacy preservation capabilities of audio sensing systems, we propose a multi-faceted approach that encompasses both speech and speaker leakage. For speech leakage, we use recent ASR models to understand the leakage of words and phonemes. Additionally, we also evaluate speaker attributes leakage across the existing techniques. For speaker leakage, we identify important speaker demographic attributes such as age, gender, and ethnicity and assess their leakage from existing privacy preserving techniques. We specifically choose these speaker characteristics as they can reveal sensitive personal information and enable various privacy attacks. While age and gender information may result in discriminatory treatment or targeted advertising, accent or dialect features could reveal a speaker's geographic or cultural background. By measuring speaker attributes leakage, we provide a more comprehensive assessment of privacy risks and enable the development of truly privacy-preserving audio sensing systems. We use Common Voice dataset [23] that contains labeled audio and corresponding text, age, gender, and accent. We assess the efficacy of predicting these speaker attributes after applying existing privacy-preserving techniques. Speech leakage refers to how much of the speech content (words, phonemes, sentences) is still intelligible after filtering, focusing on preserving linguistic content. Speaker leakage refers to how much of the speaker-specific attributes (e.g., pitch, formants, intonation, spectral energy) are preserved, focusing on identifying the individual speaker or their traits like age, gender, and ethnicity).

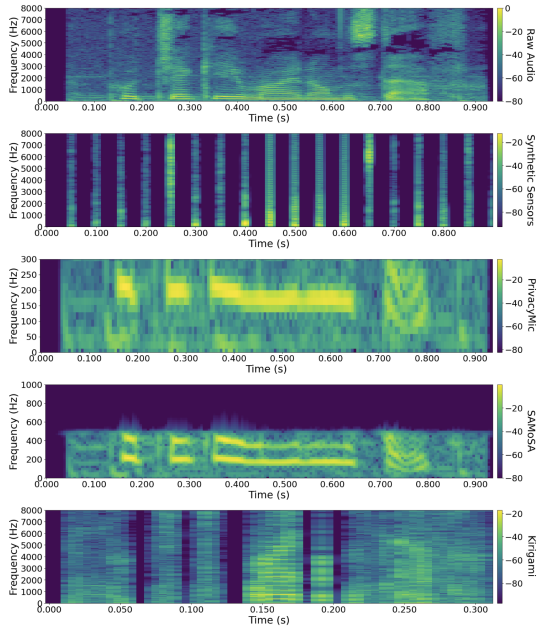

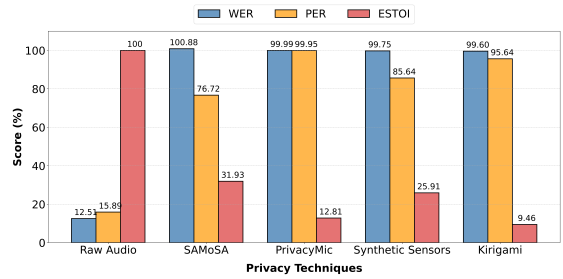

Table 2. Comparison of Privacy Techniques

Fig. 5. Speech Leakage across Privacy Techniques: WER, PER and eSTOI comparison

Fig. 4. Spectrograms of Existing Privacy Techniques

**Existing Techniques**: We evaluate several prominent privacy-preserving audio sensing techniques to assess their effectiveness in protecting both speech content and speaker information. Our analysis focused on Kirigami [11], SAMoSA [16], PrivacyMic [29], and Synthetic Sensors [30]. We specifically choose these techniques as they use different privacy-preserving techniques. SAMoSA operates in the time domain, subsampling audio at lower rates (≤ 1 KHz) to remove fine-grained speech details but potentially preserving low-frequency components that may still carry intelligible speech or speaker identity. Synthetic Sensors reduce the resolution of Fast Fourier Transform (FFT) data and sample 10 times per second, working in the frequency domain to obscure speech patterns, though overlapping windows and retained frequency bands might enable partial reconstruction of speech features like pitch or formants. PrivacyMic applies a low-pass filter to remove frequencies above 300 Hz, aiming to limit speech intelligibility; however, frequencies within the typical speech band (20-300 Hz) may still allow partial reconstruction of phonemes or speaker characteristics. While these methods strive to balance privacy and utility, they carry risks of speech leakage [11]. Kirigami uses a Logistic Regression model to eliminate speech segments from the audio signal. Although, speech can be aggressively eliminated by tuning the privacy-utility trade-off parameter, filtered signal might retain speaker-specific features. For Kirigami, we use the pretrained phoneme filter model with threshold value of 0.5.

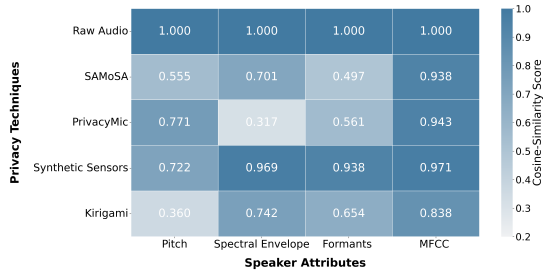

Fig. 6. Leakage of speaker-specific features across privacy-preserving filters

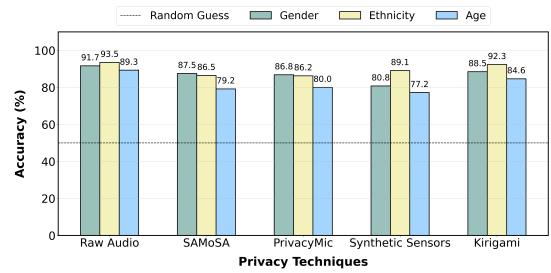

Fig. 7. Speaker Demographic Leakage across privacy techniques

Figure 4 shows spectrograms after using these privacy techniques. It can be seen from the spectrograms that the filtered spectrograms still contain a significant amount of spectral and temporal information. The key components that could leak speech or speaker information include formants (vowel and phoneme structures) and harmonics (speaker-specific traits like pitch). The fundamental frequency (pitch), retained in low-frequency regions of PrivacyMic and partially in Synthetic Sensors, can reveal speaker traits like gender and identity. Prosodic features (rhythm, intensity, and intonation) are preserved as periodic patterns in SAMoSA and Synthetic Sensors, which may leak speaker characteristics. These components, even when partially retained, can compromise privacy.

### 4.1 Speech Leakage

We measure speech leakage using three metrics: WER, Phoeme Error Rate (PER), and Extended Short-Time Objective Intelligibility (eSTOI). WER measures the distance between the recognized text and the reference transcription. PER evaluates the system's ability to protect phonetic information. eSTOI assesses the intelligibility of the processed audio. Lower values of WER and PER indicate more speech leakage, while higher values of eSTOI indicate more speech leakage. Figure 5 shows speech privacy evaluation through WER, PER and eSTOI metrics. These metrics are derived using Facebook's pretrained Wav2Vec2 [47] model on TIMIT dataset [48]. The system utilizes the Wav2Vec2 model for speech recognition, backed by phoneme mappings to analyze phoneme-level accuracy. Audio files from the dataset are loaded, normalized, and processed using methods such as SAMoSA (downsampling), Synthetic Sensors (reducing spectral resolution via STFT, and reconstruction), Kirigami (speech removal) and PrivacyMic (applying low-pass frequency filtering. Each file's transcription accuracy is evaluated using WER and PER, comparing predicted transcriptions and phoneme sequences against the ground truth. Intelligibility is assessed with eSTOI by comparing raw and processed audio signals.

Figure 5 shows speech leakage performance metrics for the privacy techniques. Raw Audio retains all spectral and temporal information, resulting in high intelligibility (100% eSTOI) and low error rates (WER/PER). In contrast, privacy techniques severely degrade intelligibility by aggressively filtering and increase error rates. Although the WER are mostly high (near 100%), there are still some phonemes being leaked by SAMoSA and Synthetic Sensors techniques as indicated by the lower PER and high eSTOI. PrivacyMic shows high PER and low intelligibility as it removes audible frequencies. These results are consistent with the results presented in Kirigami [11]. Overall, Kirigami shows the least speech leakage as it is specifically designed to prevent speech leakage with low eSTOI of 9.5%. Notably, even an eSTOI of around 10% suggests that certain speech structures remain perceptible.

## 4.2 Feasibility of Retaining Speaker-Specific Features

To understand the feasibility of retaining speaker information, we extract speaker-specific features (pitch, formants, spectral envelop, and MFCC) from both original and processed signals to check whether these features are preserved after applying privacy techniques. Pitch represents the fundamental frequency of speech, which is unique to an individual's voice due to differences in vocal cord vibration. Spectral envelope captures the overall shape of the frequency spectrum, reflecting vocal tract resonances that are specific to each speaker's anatomy. Formant frequencies represent peaks in the spectral envelope (F1, F2, F3, etc.), directly tied to the vocal tract configuration, crucial for both speech intelligibility and speaker identification. We also use MFCC features as they have been known to be the most popular, successful and widely used approach for Speaker identification [49]. We compute cosine similarity scores between raw and processed features for Common Voice [23] audio files. We analyze audio in short overlapping frames (300 ms) to capture these features over time. The formant calculation finds the first 3 peaks in the frequency spectrum using methods Linear Predictive Coding (LPC). For MFCC, we extract 20 MFCC coefficients after sliding windowing.

Figure 6 shows the similarity metrics for pitch, spectral envelope, formant frequencies, and MFCCs across different privacy-preserving filters. Raw audio serves as a baseline, with perfect similarity scores of 1.0 for all metrics. SAMoSA retains MFCCs (0.938) well but significantly reduces formant similarity (0.497). PrivacyMic preserves pitch (0.771) but heavily distorts the spectral envelope (0.317), suggesting it affects timbral qualities. Synthetic Sensors maintains high similarity in spectral envelope (0.969), formants (0.938), and MFCCs (0.971), indicating minimal impact on speaker attributes. Kirigami, designed for speech elimination, shows the lowest similarity in pitch (0.360) but moderately retains spectral envelope (0.742), formants (0.654), and MFCCs (0.838), demonstrating its effectiveness in reducing identifiable speaker traits while preserving useful non-speech features. These results highlight that while some filters are more effective than others, none fully eliminate speaker-identifying features.

## 4.3 Speaker Demographic Leakage

To evaluate speaker-related privacy leakage, we utilize the Mozilla Common Voice dataset containing speech samples with demographic attributes like age, gender, and ethnicity. We extract standard acoustic features (Mel Spectrogram) from both raw and privacy-processed audio, which are then used to train Random Forest classifiers for each demographic attribute. We use 5000 files from Common Voice dataset for this experiment and train binary classifiers for gender (Male/Female), age (Teens/ Seventies), and ethnicity (US/Indian). The results for all the files and multi-class classifiers are show in the Section 8.3. These classifiers attempt to predict speaker characteristics from the processed audio, with classification accuracy serving as a metric for privacy leakage. Higher accuracy indicates that more demographic information remains detectable despite privacy processing, while lower accuracy suggests better protection of speaker identity. This quantitative approach allows us to directly measure how effectively each privacy technique obscures sensitive demographic information.

Figure 7 shows the speaker attribute leakage of different privacy-preserving filtering approaches. Raw audio shows the highest accuracy for detecting attributes like gender, accent, and age, indicating maximum information leakage. As discussed in the earlier, all these techniques still retain speaker-specific features, causing leakage. Kirigami shows maximum leakage across all attributes as we use a privacy-utility threshold of 0.5. SAMoSA and Synthetic Sensor also show high leakage speaker-related attributes, as it retains high accuracy across attributes.

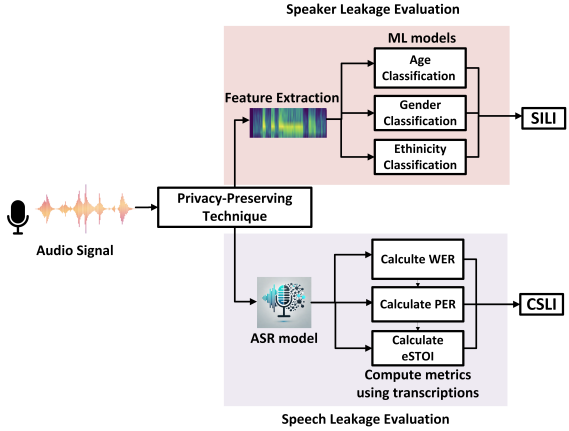
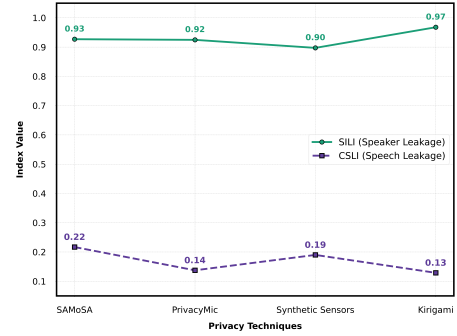

Fig. 8. SILI AND CSLI computation

Fig. 9. Speech Leakage Vs Speaker Leakage

PrivacyMic achieves relatively low leakage due to its low-pass filtering approach. The speaker leakage trends are influenced by the type of filtering applied and its impact on speaker-specific features such as pitch, formants, and harmonics. SAMoSA, which uses time-domain subsampling, retains much of the original signal's temporal structure, resulting in higher leakage of speaker identity attributes. PrivacyMic and Synthetic Sensors, which apply frequency-domain filters like low-pass filtering and reduced FFT, effectively distort higher-frequency components still carry speaker-specific characteristics.

These results demonstrate that while existing techniques offer varying levels of privacy protection, they still leak substantial speaker information. That is, our evaluation using the Common Voice dataset shows that even with current privacy methods, we can still extract sensitive attributes like age, gender, and ethnicity from the processed audio features with concerning accuracy. These results indicate that simply filtering or degrading audio or removing speech segments is insufficient for ensuring speaker privacy. *This motivates the need for more robust privacy-preserving techniques that can maintain activity/event recognition performance while more effectively obscuring sensitive speaker attributes.* Our findings underscore the need for more robust privacy-preserving techniques that address both speech content and speaker attribute preservation. In the following sections, we will introduce our novel approach that significantly reduces these privacy leakages while maintaining high utility for various audio sensing tasks.

### 4.4 Assessing Speech and Speaker Leakage: SILI and CSLI

We propose a new metric called Speaker Information Leakage Index (SILI) that combines multiple dimensions of speaker-related information leakage (e.g., gender, age, ethnicity) into a unified metric. It quantifies leakage as the average percentage of retained speaker information across these dimensions.

$$\text{SILI} = \sum_{i=1}^{N} w_i \cdot \frac{A_i}{A_{\max}} \tag{1}$$

where $N$ is the total number of speaker attributes (e.g., gender, age, ethnicity). $A_i$ is the classification accuracy for the $i$-th attribute after applying the privacy-preserving technique. $A_{\max}$ is the maximum achievable accuracy (baseline accuracy without applying privacy-preserving techniques). $w_i$ is the weight assigned to the $i$-th attribute, reflecting its importance (default: equal weights, $w_i = \frac{1}{N}$). The steps to compute SILI involve selecting attributes of interest, such as gender, age, or ethnicity, that represent the speaker characteristics being assessed. Next, measure the baseline accuracy ($A_{\max}$) by determining the classification accuracy using raw data. Following this, compute the privacy-preserved accuracy ($A_i$) by evaluating classifiers on data processed by the privacy-preserving technique. Then, apply weights ($w_i$) to each attribute based on its relevance. We use equal weights for each attribute. Finally, compute SILI by calculating the average normalized accuracy across all attributes. The SILI metric provides a unified and interpretable measure of privacy leakage, enabling researchers to compare privacy-preserving techniques. A higher SILI value indicates lesser privacy preservation. SILI ranges from 0 to 1, where 1 indicates maximum leakage.

**Advantages of SILI over EER.** While EER is a standard metric in biometric systems for assessing the trade-off between false acceptance and false rejection rates, it primarily focuses on identity verification tasks and requires a database of user-specific signatures. This requirement poses challenges in privacy-preserving scenarios where storing identifiable user data is undesirable. In contrast, SILI offers a more nuanced evaluation of privacy by quantifying the extent to which speaker identity information can be inferred from audio features *without requiring user signatures*. SILI provides a *granular assessment of privacy risks*, enabling developers to identify and mitigate specific vulnerabilities in audio representations. Moreover, SILI is adaptable to various applications such as emotion detection, offering a comprehensive framework for evaluating user privacy. For example, we can assign higher weights to the attributes that should be private for a given application. Furthermore, we can add more attributes to this metric, making it expandable. We propose developing standardized attribute sets and weighting guidelines tailored to specific applications and threat models. Additionally, introducing sub-indexing within SILI can provide granular insights into individual attribute leakage, facilitating clearer interpretation.

**Composite Speech Leakage Index (CSLI)**: Our Composite Speech Leakage Index (CSLI) metric quantifies the amount of intelligible speech information leaked by a privacy-preserving method, relative to raw audio. It combines three key metrics: WER, PER, eSTOI. The CSLI is defined as:

$$\text{CSLI} = w_1 \cdot \frac{\text{WER}_{\text{raw}}}{\text{WER}_{\text{method}}} + w_2 \cdot \frac{\text{PER}_{\text{raw}}}{\text{PER}_{\text{method}}} + w_3 \cdot \frac{\text{eSTOI}_{\text{method}}}{\text{eSTOI}_{\text{raw}}} \quad (2)$$

Here, $w_1$, $w_2$, and $w_3$ are weights assigned to each metric, with $w_1 + w_2 + w_3 = 1$. The terms $\text{WER}_{\text{raw}}$, $\text{PER}_{\text{raw}}$, and $\text{ESTOI}_{\text{raw}}$ represent the baseline scores for raw audio, while $\text{WER}_{\text{method}}$, $\text{PER}_{\text{method}}$, and $\text{ESTOI}_{\text{method}}$ are the corresponding scores for the privacy-preserving method being evaluated. Higher values of WER and PER of the privacy-technique means less speech leakage, leading to lower CSLI values. Higher values of eSTOI indicates high leakage, leading to high CSLI. Thus, a higher CSLI value indicates greater speech information leakage, with raw audio typically serving as the maximum leakage baseline.

Figure 9 highlights key differences in speaker and speech leakage across methods. Among privacy-preserving methods, Kirigami achieves lowest speech leakage (CSLI) but highest speaker leakage (SILI). PrivacyMic also achieves lower CSLI and high SILI, indicating it is effective at protecting speech, but not speaker leakage. SAMoSA and Synthetic Sensors leak high speech content and similar speaker leakage as PrivacyMic. An important observation is that speaker leakage (SILI) and speech leakage (CSLI) are not the same. Techniques that do not leak speech content might still leak speaker information.

**Key Takeaways**: Speech leakage is distinct from speaker leakage, and existing privacy-preserving techniques leak significant speaker information. This motivates the need for new techniques to prevent speaker leakage. We present SILI as a new metric capture the amount of speaker leakage in any audio sensing task.

## 5 FEATURESENSE OVERVIEW

### 5.1 Overall Approach

*FeatureSense* uses a feature-based audio approach to prevent speaker leakage. To preserve user privacy, our approach extracts a set of privacy-preserving audio features from the raw audio signal in real-time, as shown in

Figure 3. The approach then exposes these features, rather than the raw audio, for downstream tasks, thereby to enable a range of audio sensing applications while preserving privacy. *We hypothesize that by using carefully designed granular audio features, we can preserve speaker privacy and still perform common audio classification tasks (non-speech, speaker-invariant).* Our feature-based approach is suitable for a wide range of audio classification tasks that do not include speech, such as environment sound classification. Additionally, we propose tunable parameters that maintain privacy-utility-cost trade-offs by dynamically selecting subsets of features for the desired application. Through this work, we explore the following research questions: a) What are the privacy-preserving and non-privacy preserving audio features?, b) What all can we do with privacy-aware audio features?, and c) What is the computational overhead of extracting these features? By answering these questions, we aim to assist developers in designing different audio sensing applications.

## 5.2 Overview of Audio Feature Groups and Their Privacy Properties

**Non-Privacy-Invasive (Green), Grey Zone (Yellow), Privacy-Invasive (Red)**

| Time Domain | Spectral | Phonetic/ Linguistic | Statistical | Perceptual | High-Level | Voice-Specific | Derived |
|---|---|---|---|---|---|---|---|
| Amplitude Envelope | Spectral Centroid | Formants | Mean | Sharpness | Group Delay | Pitch | Low Band Energy |
| RMS | Spectral Flatness | Filter Bank | Variance | Timbre | Wavelet Features | HNR | Mid Band Energy |
| Zero Crossing Rate (ZCR) | Spectral Contrast | LPCC | Standard Deviation | Reverberation | Temporal Spectral Slope | Jitter | High Band Energy |
| Short-Term Energy (STE) | Spectral Spread | Fundamental Frequency | Kurtosis | Tonality Index | Chroma Features | Shimmer | Low-to-High Ratio (LH1000) |
| Temporal Centroid | Spectral Irregularity | Spectral Envelop | Skewness | | Group Delay | | Spectral Texture |
| Envelope Modulation Rate | Spectral Entropy | | | | | | Spectral Roughness |
| Silence Ratio | Spectral Roughness | | | | | | Transient-to-Sustained Ratio |

Table 4. Privacy Sensitivity of Various Audio Features

We use categorize audio features into 8 major classes: *Time-Domain, Spectral, Phonetic, Statistical, Perceptual, High-level, Voice-specific, and derived features* as shown in Table 4.

**Time-domain features** are derived from the raw amplitude variations of an audio signal over time. Key features in this category include Amplitude Envelope which tracks the variations in loudness, Root Mean Square (RMS) which measures the average power of the signal), Zero Crossing Rate (ZCR) which counts the rate at which the signal crosses the zero amplitude line (useful in distinguishing voiced and unvoiced speech). Features like Short-Term Energy (STE) are used to detect high-intensity events or transitions in audio. Advanced metrics such as Temporal Centroid and Envelope Modulation Rate describe the timing characteristics of energy distribution and rhythmic patterns. Envelope Modulation Rate measures the average rate of change in the amplitude envelope of the signal while Temporal Centroid is the "center of mass" of the signal's energy in the time domain. Silence Ratio is a measure of the proportion of silence in the signal. These features are generally non-privacy-invasive since they capture general signal properties rather than detailed personal or linguistic information.

**Spectral features** are extracted from the frequency representation of an audio signal, providing insights into its frequency content and distribution. Spectral Centroid indicates the "center of mass" of the frequencies and is often used to characterize the brightness of sound. Spectral Flatness measures how noise-like a sound is, which is useful for distinguishing between harmonic and percussive sounds. Spectral Spread describes the range of frequencies present in a signal, while Spectral Contrast highlights the difference between peaks and valleys in the spectrum. Features such as Spectral Entropy, Spectral Irregularity provide detailed statistical characterizations of the frequency distribution.

**Phonetic and linguistic features** directly analyze speech content and speaker characteristics. Examples include Fundamental Frequency (F0), which corresponds to the pitch of speech and is used in gender and emotion recognition. Mel-Frequency Cepstral Coefficients (MFCCs) are commonly used in automatic speech recognition and speaker identification due to their ability to represent the phonetic content of speech [49]. Other features such as Formant Frequencies, Harmonic Ratios, Spectral Envelop and Filter Banks capture detailed phonetic characteristics and vocal tract resonances, making them useful for speaker and language identification. Linear Predictive Coding (LPC) models the vocal tract, making it central to speech synthesis and recognition. These features are often privacy-invasive, as they can reveal not only the speech content but also personal attributes like identity, accent, etc.

**Statistical features** provide descriptive statistics of the audio signal, such as Mean, Variance, Standard Deviation, Kurtosis, and Skewness. These features summarize the overall distribution of amplitude or frequency content. Entropy, which measures the randomness of the signal, is particularly useful in detecting structured versus unstructured sounds. These features are generally non-privacy-invasive as they do not capture specific personal or linguistic characteristics.

**Perceptual features** are designed to model how humans perceive sound, often derived from psychoacoustic principles. Sharpness measures the perceived brightness of sound, while Timbre captures the texture or color of the sound. Reverberation quantifies the presence of echoes, often used in room acoustics analysis. Tonality Index measures the ratio of harmonic (tonal) energy to noisy (aperiodic) energy in a signal, quantifying how tonal or noise-like the sound is. While most perceptual features are non-privacy-invasive, Timbre can be privacy-invasive as it can encode speaker-specific characteristics.

**High-level (Time-frequency) features**, such as Temporal Spectral Slope, Group Delay, and Wavelet Features, Chroma features provide advanced representations of the signal by combining time and frequency information. These features are particularly useful in tasks like sound event detection, environmental audio classification, and advanced signal processing. Temporal Spectral Slope measures the rate of change of spectral centroid over time, Group Delay represents the time delay of the signal's frequency components, useful for analyzing phase characteristics and resonances. Wavelet Features extract multi-scale time-frequency information using wavelet transformations, capturing transient and localized patterns in audio. Chroma Features represent the energy distribution across pitch classes. High-level features are typically non-privacy-invasive as they focus on abstract signal properties rather than personal or linguistic details. However, Chroma Features can fall into a gray zone due to their potential to identify specific audio content or speakers.

**Voice-specific features** are tailored to analyze speaker characteristics and voice quality. Harmonics-to-Noise Ratio (HNR) evaluates the amount of harmonic content versus noise in a voice, often used in clinical speech assessments. Jitter and Shimmer measure variations in pitch and amplitude, respectively, providing insights into voice quality. These features fall into the gray zone of privacy, as they can encode speaker-specific attributes without directly revealing speech content.

**Derived Features**: Additionally, we explore some derived features like low band energy (energy below 500 Hz), Mid band energy (energy between 500-2000 Hz), high band energy (energy above 2000 Hz), and Low-to-High ratio (LH1000) which is the ratio of energy below 1000 Hz to the energy above 1000 Hz. We use these features because different frequency bands are relevant for different environment sound classes. Like Low-frequency sounds are often dominant in environmental sounds like engines, machinery, etc. High-frequency content is common in sounds like bird chirps, glass breaking, or alarms, making this feature important for distinguishing such events. We also use Transient-to-Sustained Ratio, which is the ratio of energy in transient events (short, sharp sounds) to sustained energy (continuous sounds). Transients are characteristic of sounds like knocks or crashes, whereas sustained energy represents continuous sources like wind or engine noise. This ratio helps differentiate between such sound types. Spectral Roughness quantifies the perceptual dissonance or irregularity between neighboring harmonic components in the spectrum. It helps in identifying complex or rough textures in sounds like rustling leaves or industrial noise. Spectral Texture Coefficients measures the multi-scale variability of the spectral energy distribution using wavelet analysis. It captures the texture of environmental sounds, which is vital for differentiating between smooth sounds (e.g., rain) and more chaotic ones (e.g., crowd noise). These features are designed to capture non-speaker-specific characteristics of audio signals, focusing on environmental and general sound properties rather than individual or linguistic traits.

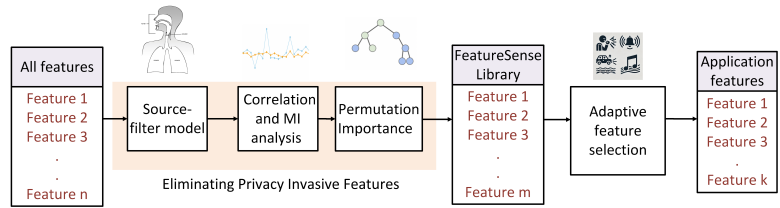

Fig. 10. Steps to Eliminate Privacy-Invasive Features

### 5.3 Challenges in Feature-Based Audio Classification

Feature-based approaches offer significant advantages in efficiency and interpretability, but present several non-trivial challenges. First, **preventing speaker attribute leakage** is difficult, as even low-level acoustic features (e.g., formants, spectral descriptors, MFCCs) can inadvertently encode demographic traits such as age, gender, or ethnicity. Most existing feature-based systems do not explicitly evaluate such leakage. We identify and eliminate the features that can uncover these speaker attributes using statistical and ML-driven analysis (Section 6.1). Second, while feature-based systems avoid deep models, **feature extraction itself can introduce latency**, especially when computing a large, general-purpose set of features. This introduces a new cost-privacy-utility tradeoff—distinct from traditional model latency—that must be managed carefully in real-time applications (Section 6.3). Third, designing effective and privacy-conscious feature sets often requires **domain expertise**, limiting the accessibility of these methods to non-experts. Moreover, different applications (e.g., health monitoring vs. smart homes) demand different balances of privacy, utility, and latency. To address these challenges and enable context-awareness, *FeatureSense* introduces an adaptive feature selection algorithm that automatically selects an optimal subset of features given task-specific constraints (Section 6.3). This enables developers to build privacy-aware audio systems without requiring deep acoustic knowledge.

## 6 FEATURESENSE DESIGN

In this section, we present the design of our *FeatureSense* system.

### 6.1 Feature Selection to Prevent Speaker Leakage

**Eliminating privacy-invasive features**: The *source-filter model* of speech production describes speech as the output of two primary components: a source and a filter. The source represents the airflow and vocal fold vibrations that generate the fundamental frequency (pitch) and harmonic structure, while the filter corresponds to the resonant properties of the vocal tract, shaping the spectral envelope of the speech signal. In this model, certain acoustic features are strongly linked to speech content (what is being said) and speaker identity (who is speaking). Features like formants (resonant frequencies shaped by the vocal tract) are crucial for speech intelligibility and linguistic content, as they define vowels and phonemes. On the other hand, features such as fundamental frequency and spectral envelop closely tied to speaker-specific characteristics, as they reflect the physiological properties of the speaker's vocal folds and vocal tract [50]. The vocal tract resonances, called formants (especially F1, F2, and F3), are particularly vital as they help differentiate and recognize vowels, which are crucial for intelligibility. If we do not capture the first three formants, it is hard to reconstruct speech. LPCCs primarily capture vocal tract filter properties, making them relevant to both speech content (formants, phoneme identification) and speaker identity (vocal tract shape, resonance patterns). So, we eliminate these phonetic/linguistic features that are known to leak privacy. We only explore features that are not privacy-invasive or lie in the gray zone as shown in the Table 4.

**Privacy Leakage of Individual Features**: We measure contribution of individual feature in leaking important speaker attribute. Correlation measures linear relationships between features and attributes. However, privacy leakage could arise from non-linear or complex relationships. So, we use mutual information which measures both linear and non-linear dependencies between features and sensitive attributes. Higher mutual information or correlation indicates stronger predictive power. Figure 11 shows the correlation and mutual information for all the features with gender. Features like Spectral Contrast, Chroma features, Wavelet features and Timbre show relatively high correlation, but it is still less than 0.3. Amplitude Envelop and Chroma features show high mutual information with gender. Additionally Timbre and Wavelet features show relatively high MI score. We use Random Forest model to measure feature importance. Additionally, we use Permutation Importance (PI) to measure model-agnostic importance from feature perturbations. Since Timbre and Chroma Features show highest high values across all metrics, indicating potential speaker attributes leakage. We eliminate these features from the *FeatureSense* library. In Section 8.3, we show that the removal of these features can significant reduce the accuracy of the age, gender, and ethnicity classification models, thus preserving privacy. We perform a similar analysis for age and accent features as well. For age, all features showed <0.1 correlation with age labels. For accent, all features showed <0.04 correlation with accent labels. Similar values were observed for MI and PI, indicating that the features do not show predictive ability for age and accent.

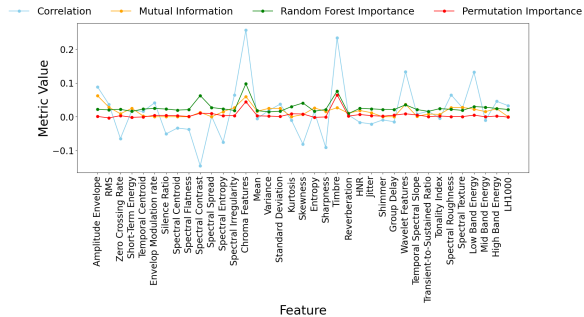

Fig. 11. Correlation, MI, Permutation Importance and Random Forest Importance of individual features with gender

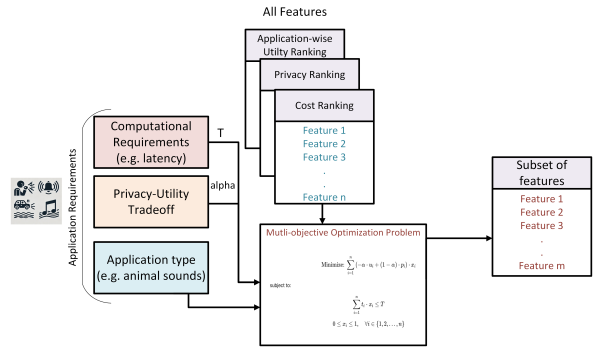

Fig. 12. Task-Specific Feature Selection Overview

## 6.2 FeatureSense Library: Privacy-Aware Audio Features for Non-speech Speaker-invariant Sensing

Our *FeatureSense* library provides a set of features which are granular, privacy-aware, general-purpose, and useful for a wide range of audio classification tasks. Each feature represents a distinct and interpretable property of the signal. We chose these features carefully chosen for their relevance to specific downstream tasks like environment sound classification, non-speech human sound detection, etc. These features contain less sensitive information than high-dimensional representations, reducing privacy risks. We use time-domain, spectral, perceptual, voice-specific, statistical, time-frequency, and derived features. We do not use phonetic/linguistic features as they are known to leak speech and speaker related information. The feature list includes *Amplitude Envelope, RMS, Zero Crossing Rate, Short-Term Energy, Temporal Centroid, Envelop Modulation rate, Silence Ratio, Spectral Centroid, Spectral Flatness, Spectral Contrast, Spectral Spread, Spectral Entropy, Spectral Irregularity, Mean, Variance, Standard Deviation, Kurtosis, Skewness, Entropy, Sharpness, Reverberation, HNR, Jitter, Shimmer, Group Delay, Wavelet Features, Temporal Spectral Slope, Transient-to-Sustained Ratio, Tonality Index, Spectral Roughness, Spectral Texture, Low Band Energy, Mid Band Energy, High Band Energy, and LH1000 as shown in green in the Table 4*. Instead of exposing raw audio data, which contains sensitive information that could be used for unintended purposes (e.g., speaker identification or tracking), we extract and expose a curated set of privacy-aware features which can be useful for many non-speech speaker-invariant tasks. These features encapsulate critical information needed for tasks such as environment noise detection, smoke alarm detection, cough detection, etc, while mitigating the risk of disclosing personally identifiable information. Our library takes audio data from the microphone as input and outputs a list of these computed audio features. We also provide an Application Programming Interfaces (APIs) to configure parameters like window size, feature groups, etc. Our approach is based on allowing users to select and work with low-dimensional features instead of raw audio. This enables the use of audio data in applications without the risk of exposing sensitive, speaker-specific information. One key element of our system is the ability to evaluate the privacy implications, utility, and cost of individual features. We make these scores accessible through library APIs. Privacy rankings are computed from the Gini importance scores derived from age, gender, and ethnicity classifiers and utility rankings are derived from the Gini importance scores derived from the ESC-50 classifier as discussed in Section 6.3. We characterize the computational cost of each features as the latency to extract it using an optimized implementation that uses low level Python libraries. Our library is modular, extensible, and equipped with interfaces to optimize feature subsets based on task requirements, context, privacy concerns, computational constraints, and utility metrics (as detailed in Section 7), to support informed decision-making.

## 6.3 Task Specific Feature Selection

**Privacy-Utility-Cost Trade-offs**: We focus on computing the trade-offs of including each feature for a given application. For each feature, we compute a privacy, utility, and the computation cost. To compute the privacy metric, we use the gini importance of the feature for age, gender, and accent classification models trained on Common Voice dataset [23]. The higher the metric, the more it leaks privacy. To compute utility metric, we use the gini importance score of the feature on ESC-50 [26]. Since ESC-50 has 5 main sound categories (animal sounds, non-speech human sounds, interior sounds, urban sounds, and nature sounds), we have separate utility score of the features for each sound category. The higher score indicates high utility of the feature. To compute the cost metric, we use the feature extraction latency of the feature. The higher values indicate higher cost. Having the privacy leakage scores, application-specific utility scores, and the cost scores for all the features, we design a linear programming based algorithm to compute the optimal feature list for a given application. Designing audio classification systems requires selecting features that balance task-specific requirements for utility (classification accuracy), computational efficiency, and privacy preservation. This requires manual effort of feature engineering and domain knowledge. We propose a multi-objective optimization framework to automatically select the optimal subset of features for a given application as shown in Figure 12. First, we select the category (animal, human, nature, interior, and exterior sounds) of application and use its utility scores. Next, the feature selection process is framed as a multi-objective optimization problem with three primary objectives:

(1) **Maximizing Utility:** Features should contribute significantly to the classification performance, as measured by feature importance or accuracy on the target task.

(2) **Minimizing Computational Cost:** Feature extraction should meet latency or resource constraints, particularly for edge devices or real-time applications.
(3) **Minimizing Privacy Leakage:** Features should avoid encoding sensitive information to ensure compliance with strict privacy requirements.

The algorithm combines the objectives of maximizing utility and minimizing privacy leakage, controlled by a single trade-off parameter, $\alpha$. This parameter allows the user to prioritize utility or privacy leakage based on the application requirements. The optimization objective is defined mathematically as:

$$\text{Objective: } \sum_{i=1}^{n} (\alpha \cdot u_i - (1 - \alpha) \cdot p_i) \cdot x_i,$$

where $x_i$ is the binary decision variable indicating whether the $i$-th feature is selected ($x_i = 1$) or not ($x_i = 0$). The term $u_i$ represents the utility of the $i$-th feature for a specific application category, while $p_i$ denotes the average privacy leakage of the feature, calculated as the mean of privacy leakage values across multiple dimensions such as age, gender, and ethnicity. The parameter $\alpha$ controls the trade-off: when $\alpha = 1$, the algorithm focuses entirely on maximizing utility; when $\alpha = 0$, it prioritizes minimizing privacy leakage. The optimization is subject to the following constraints:

*Latency Constraint*: The total latency of the selected features must not exceed a predefined threshold $T$: $\sum_{i=1}^{n} t_i \cdot x_i \leq T$, where $t_i$ represents the latency of the $i$-th feature.

*Binary Selection Constraint*: Each feature is either selected ($x_i = 1$) or not ($x_i = 0$): $0 \leq x_i \leq 1, \quad \forall i \in \{1, 2, \ldots, n\}$.

The proposed algorithm loads the utility, privacy, and latency data for all features. This data is merged into a single dataset, ensuring the inclusion of application-specific utility scores and privacy leakage values. The coefficients of the objective function are computed as $c_i = -\alpha \cdot u_i + (1 - \alpha) \cdot p_i$, which balances utility and privacy leakage. The constraints are then formulated and a linear programming approach is used to minimize the objective function under the specified constraints. Finally, the features corresponding to $x_i = 1$ are extracted as the selected subset. This algorithm is designed to address the inherent trade-offs in feature selection by providing flexibility through the parameter $\alpha$. The design reflects a practical need in applications like audio analysis, where features can serve multiple goals, such as maximizing classification performance while reducing the risk of privacy leakage. By combining utility and privacy into a single objective, the algorithm ensures a holistic approach to feature selection. The latency constraint adds an additional layer of practicality to the algorithm, ensuring that the selected features can be computed efficiently within real-time or application-specific limits. This makes the algorithm particularly suitable for resource-constrained environments, such as mobile or embedded systems, where both computational overhead and privacy risks need to be managed. Increasing $\alpha$ prioritizes utility at the cost of privacy, while decreasing $\alpha$ does the opposite. This simplicity in design allows the algorithm to be easily understood and adapted across various domains and use cases. Ultimately, this algorithm exemplifies a balance between performance optimization and ethical considerations, such as preserving user privacy.

Our framework enables the design of audio classification systems tailored to specific applications. In utility-driven applications, we can prioritize features with the highest utility. For privacy-sensitive applications, such as voice assistants, avoid features with high privacy leakage. For latency-constrained applications, computationally efficient features can be preferred while maintaining sufficient utility. The multi-objective optimization framework provides a systematic and automated approach to feature selection, allowing practitioners to design audio classification systems that are optimized for their specific requirements without requiring domain knowledge. By balancing utility, computational efficiency, and privacy, the framework ensures that the selected features meet both performance and ethical considerations.

## 7 FEATURESENSE IMPLEMENTATION

Our system takes raw audio input from the microphone and processes it into a series of privacy-aware features. The key component is a feature extraction module, which processes the signal, removing raw speaker identity-related details. We design a modular and extensible audio feature library (similar to Opensmile [51]) that provides an integrated framework for feature extraction, accessing privacy-utility-cost metrics of the features, and adaptive feature selection based on application requirements. Our library is structured into two primary components: the feature extraction module and the task-specific feature selection module. The **feature extraction module** takes the microphone device ID as input and returns dictionary of audio features in real-time. It includes a wide range of audio features, divided into 7 logical categories. These categories include time-domain features (e.g., RMS, Zero Crossing Rate), spectral features (e.g., Spectral Centroid, Spectral Flatness), perceptual features (e.g., Timbre, Sharpness), voice-specific features (e.g., jitter, HNR), statistical (e.g., mean, variance), high-level time-frequency features (e.g., Group Delay, Wavelet Features), and derived features (e.g., Low band energy, transient-to-sustained ratio). To ensure robust handling of non-stationary signals, we use a sliding window approach, which processes audio signals in overlapping segments, enabling localized feature extraction. The window size and sampling rate can be set as passed as parameters (default value: 500 ms and 16 kHz). If we need selective features, we can pass the feature list as an argument to the feature extraction module. Computational efficiency is achieved through the use of libraries such as Scipy and Numpy and custom-implemented methods for advanced features like reverberation estimation. Additionally, we can access three metrics for each feature: utility, privacy, and computational cost. As we already have privacy, utility, and cost ranking of each feature for the task-specific feature selection (as discussed in Section 6.3), we provide access to it using the APIs. Utility reflects the relevance of a feature to a target application (e.g., animal sounds, nature sounds) and is measured using gini importance score on ESC-50 dataset. Privacy leakage is assessed by estimating the extent to which a feature reveals sensitive information, such as age, gender, or ethnicity, using gini importance score of classifying these attributes. Computational cost is quantified by profiling the latency for extracting each feature.

| Function | Description | Input Parameters | Output |
|---|---|---|---|
| **Feature Extraction** `extract_privacy_features()` | Extracts selected audio features from raw audio. | `device_id` (str), `feature_list` (list,optional), `window_size` (float,optional), `sampling_rate` (int,optional) | Dict of feature values |
| **Metrics Query** `get_metrics()` | Retrieves privacy, utility, and computational cost scores. | `feature_name` (str) | Dict of scores |
| **Task-specific Feature Selection** `optimize_features_with_puc_tradeoff()` | Selects an optimal feature subset based on constraints. | `category` (str), `latency` (float,optional), `alpha` (float,optional) | Optimized feature list |

Table 5. FeatureSense API Summary

The **task-specific feature selection module** integrates the evaluated PUC metrics into a multi-objective optimization framework. It takes application category (animal sounds, nature sounds, human non-speech sounds, interior sounds, exterior sounds), latency constraint ($T$), and privacy-utility trade-off parameter ($\alpha$) as input. The default values are set to $\alpha = 0.5$, $T = 0.1$, and sound category as interior sounds. The library already has access to the CSV files containing privacy, utility, and cost rankings for all features. Using linear programming, it returns an optimal subset of features based on user-defined constraints. This list of feature can be passed to feature extraction module to extract the desired features in real time. The optimization problem is formulated to maximize utility while minimizing privacy leakage, subject to a latency constraint. The trade-off between utility and privacy is controlled using a parameter $\alpha$, where $\alpha = 1$ prioritizes utility and $\alpha = 0$ prioritizes privacy. This flexible framework allows the library to adapt to different application scenarios, such as real-time classification or privacy-sensitive analysis. We implement a user-friendly API that enables seamless interaction with the library's functionalities.

Figure 5 shows the APIs of *FeatureSense* library. The feature extraction API allows users to extract features from raw audio signals. For instance, the FeatureSense. extract_privacy_features function computes a comprehensive set of features in real time. The metrics query API provides access to precomputed utility, privacy, and cost metrics for each feature. For example, the get_metrics function enables users to retrieve privacy or utility scores for specific features, facilitating informed decision-making during feature selection. The optimization API provides task-specific feature selection capabilities. The optimize_features_with_puc_tradeoff function takes as input utility, latency, and privacy leakage data, along with parameters such as the sound category, latency constraint, and the trade-off parameter $\alpha$. It returns an optimized subset of features tailored to the user's task requirements. This linear programming framework is implemented using scipy.optimize.linprog function. This functionality simplifies the process of configuring feature sets for diverse applications. Our library uses a combination of Python libraries, including NumPy for numerical computations, SciPy for optimization, and Pandas for data handling and integration. These tools ensure high computational efficiency and scalability. This modular design, combined with intuitive APIs and scalable architecture, makes our library a powerful and flexible tool for audio analysis, addressing the needs of both research and real-world applications. Currently, *FeatureSense* is designed as a wrapper function. For *FeatureSense* deployment, we plan to have an OS-level or hardware level solution that will give access to only the privacy-aware features.

## 8 EXPERIMENTAL EVALUATION

| Dataset | Size | Classes/Labels | Used for |
| --- | --- | --- | --- |
| CommonVoice | 804 hours, 33,541 speakers | Transcriptions | Speaker Leakage |
| ESC-50 | 2,000 clips (5 sec each) | 50 classes (e.g., dog bark, siren, rain) | Utility Assessment |
| TIMIT | 5 hours, 630 speakers | Phonetic Transcriptions | Speech Leakage |
| AudioSet | 2M+ clips (10 sec each) | 600+ classes (e.g., speech, cough, sneeze) | Case Study |
| UrbanSounds8k | 8,732 clips (up to 4 sec) | 10 classes (e.g., air conditioner, car horn, drilling) | Case Study |

Table 6. Summary of Datasets Used in This Study

### 8.1 Experimental Setup

**Datasets Used**: In this study, we use four publicly available datasets to train and evaluate our models: Common Voice [23], ESC-50 [26], TIMIT [48], UrbanSounds8k [28], and AudioSet [27]. These datasets are well-known and commonly used in the literature. For instance, Kirigami [11] uses ESC-50 for utility analysis and TIMIT for understanding speech leakage. Below, we provide a detailed description of each dataset, including its characteristics, data distribution, and relevance to our research. A summary is also provided in Table 6. Common Voice is a large-scale, multilingual speech dataset provided by Mozilla. It contains speech recordings contributed by volunteers worldwide, along with their corresponding transcriptions. The dataset is designed to support automatic speech recognition (ASR) and includes diverse speakers across various accents, genders, and age groups. For this study, we utilized the English subset, comprising approximately 804 hours of speech data from 33541 speakers. ESC-50 is a curated dataset of environmental sounds, categorized into 50 classes such as animal sounds, natural sounds, human sounds, and urban noises. Each class contains 40 five-second audio clips, totaling 2,000 labeled examples. The dataset is widely used for sound classification tasks due to its balanced class distribution and clear labels. TIMIT is a phonetically transcribed speech dataset designed for speech recognition and phoneme analysis. It contains recordings of 630 speakers reading ten phonetically rich sentences, amounting to a total of 5 hours of speech. TIMIT includes both clean and noisy recordings, with detailed phonetic annotations, making it valuable for tasks such as phoneme recognition and speech synthesis. AudioSet is a comprehensive dataset of over 2 million human-labeled audio segments sourced from YouTube videos. It covers more than 600 audio event classes, including music, speech, and environmental sounds. Each audio segment is 10 seconds long and is annotated with multiple labels to reflect real-world acoustic conditions. The diversity and scale of AudioSet make it a critical resource for general-purpose audio analysis. UrbanSounds8k is a curated dataset consisting of 8,732 short audio clips (up to 4 seconds long) labeled into 10 urban sound classes, including air conditioner, car horn, drilling, dog bark, and siren. The dataset is sourced from field recordings and captures diverse real-world noise conditions commonly found in urban environments. Each clip is manually annotated and assigned to one of the predefined classes, making it a valuable resource for environmental sound classification tasks. Due to its focus on urban noise, UrbanSounds8k is widely used in machine listening, noise pollution analysis, and smart city applications. Additionally, to evaluate *FeatureSense* performance, in real world, we use a Raspberry Pi 5 based prototype connected to a USB microphones.

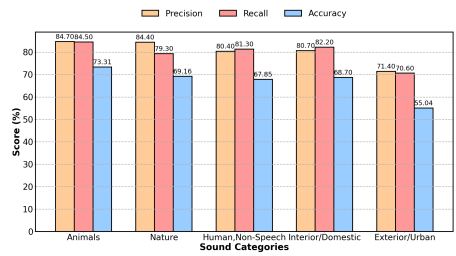

Fig. 13. FeatureSense Accuracy across Different Sound Categories

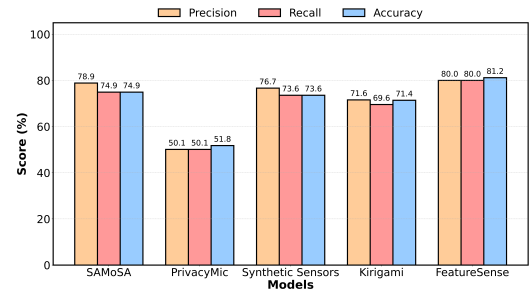

Fig. 14. Utility comparison across Different Privacy Techniques

**Metrics**: To evaluate the speaker privacy leakage, we use the proposed *SILI* metric. We compute it after applying different privacy-preserving techniques. It is computed using the classification accuracy of age, gender, accent prediction as compared to raw audio. For utility evaluation, we use classification accuracy, precision, and recall of ESC-50 as a measure. For latency evaluation, we use time in milliseconds. For speech leakage assessment, we use MI analysis with first 3 formants as at least three formants are generally required to produce intelligible speech and up to five formants to produce high quality speech [52].

## 8.2 Utility Evaluation

For assessing the utility and generalizability of the proposed approach, we use ESC-50 dataset as it has wide range of environment sounds and non-speech human sounds. With the proposed privacy-aware features, we are able to achieve average accuracy of 81.18%, average precision of 80%, and average recall of 80% using Random Forest Classifier. This is comparable to the state-of-the-art accuracy on ESC-50 using similar ML models. Figure 13 shows the classification performance across sound categories (Animals, Nature, Human Non-Speech, Interior/Domestic, and Exterior/Urban) is consistently high, with metrics like precision, recall, F1-score, and accuracy averaging above 75% for most categories. Interior/Domestic and Animals have particularly strong performance, indicating the model effectively captures features unique to these categories. However, Exterior/Urban sounds show slightly lower performance, likely due to the complexity and variability of urban noise. Overall, the model demonstrates robust classification capability but may benefit from further optimization or data augmentation in challenging categories like Exterior/Urban. Figure 14 shows that classification performances on ESC-50 after using different privacy techniques and *FeatureSense*. To evaluate the utility of existing techniques, we use the privacy filter proposed by the technique, followed by standard feature extraction (Mel Spectrogram) passed to random forest classifier. PrivacyMic shows least utility as it removes almost all audible frequencies. Other techniques show similar utility ranging between 71.4% to 74.9% aggregate classification accuracy. *FeatureSense* shows the highest utility, slightly more than SAMoSA.

**Window Size Selection**: We evaluate the impact of window size on both task utility and speaker attribute leakage using ESC-50 classification accuracy and gender detection as proxies. As shown in our experiments, shorter windows (e.g., 200 ms) yield higher task accuracy on ESC-50 (86.9%) but also higher speaker attributes leakage. In contrast, larger windows like 1 second reduce leakage significantly (by 15.7%) but degrade utility to 77.39%. We observe that a window size of 500 ms offers a balanced tradeoff, achieving 81.2% task accuracy and is therefore used throughout this paper. In future work, we plan to integrate window size into the feature selection optimization pipeline to jointly control privacy and utility (Section 9).

We calculate feature importance for the entire dataset by training a single Random Forest Classifier. Next, we train a separate Random Forest Classifier for each binary class (using LabelBinarizer) and average feature importance across classes within a category. Since this method focuses on individual classes and their unique contributions, features that are less critical for certain binary classifications may get diluted when averaged over multiple classes. Figure 15 shows the Gini importance of different feature groups across sound categories. We normalize feature importance across classes and categories, which inherently reduces the relative weight of features that are not consistently dominant in all binary classifications. Spectral features appear to be the most important across all categories, indicating their strong discriminative power for sound classification. Time Domain and High-Level features also contribute significantly, suggesting that both low-level and high-level features help in sound differentiation. Statistical and Derived features show relatively lower importance, but they still contribute to classification.

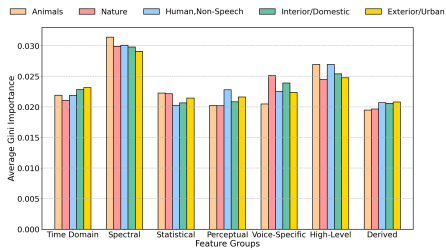

Fig. 15. Groupwise Feature Importances of Different Sound Categories

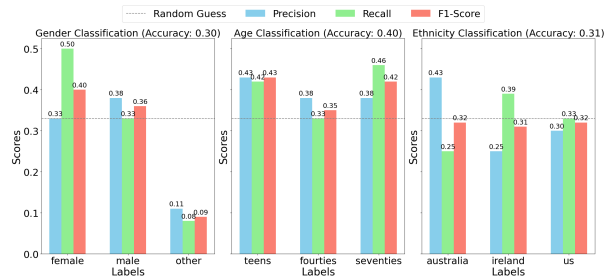

Fig. 16. Speaker Attributes Leakage with *FeatureSense*

### 8.3 Privacy Evaluation

*8.3.1 Speaker Privacy Leakage.* To assess user privacy, we use the proposed *SILI* score and compare it with the existing privacy techniques. Figure 16 shows the age, gender, and ethnicity leakage scores with our proposed approach. We use three class labels for all three attributes using Common Voice dataset: gender (male, female or other), age (teens, fourties, or seventies) and ethnicity (us, ireland, or australia). We train a classifier using all the features from the *FeatureSense* library. After removing the highest correlated features (Chroma Features and Timbre) as discussed in Section 6, we achieve close to random guess accuracy for all three speaker attributes. For example, the gender classification accuracy reduced from 48.6% to 30%, after removing the correlated features.

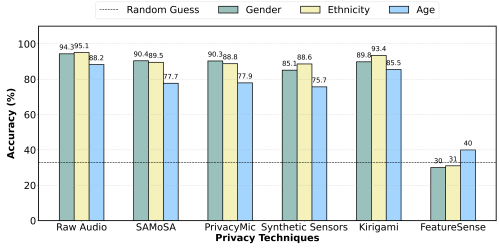

Fig. 17. Speaker Leakage Comparison across Privacy Techniques

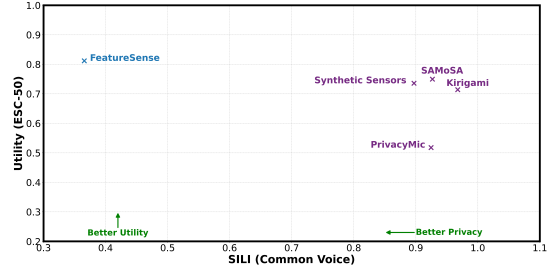

Fig. 18. Privacy-utility Trade-offs across Privacy Techniques: ESC-50 Utility Vs Speaker Privacy Leakage

The classification accuracy of speaker attributes is shown in Figure 16. The classification results show limited performance across Gender, Age, and Ethnicity tasks, with accuracies of 30%, 40%, and 31%, respectively, only marginally better than random guessing (33% for 3-class classifiers). Precision, recall, and F1-scores vary across classes, indicating imbalance in classification performance. For example, in Gender classification, the model struggles with the "other" category (F1-score: 0.09), while in Age and Ethnicity tasks, recall is higher for certain classes like "seventies" and "ireland," suggesting potential biases or insufficient feature representation for other classes. The most important features for gender classification were Kurtosis (0.063), HNR (0.044), and Spectral Spread(0.036). For age, the top features were Spectral Spread (0.053), Skewness (0.05), and Tonality Index (0.04). For ethnicity leakage, top features were Envelop Modulation rate (0.048), Tonality Index (0.042), and Skewness (0.041). Overall, the feature importance values are very low, indicating low predictive power. We compare our approach with existing privacy techniques as shown in Figure 17. It can be observed that we achieve lowest speaker leakage (close to random guess) as compared to other techniques, showing a 60.6% average improvement. Although other techniques are designed to preserve privacy by reducing audio signal information, they do measure or assess speaker-specific privacy leakage. Even after applying the sub-sampling, low pass filtering, partial sampling, and speech segment elimination, there is still significant information in the remaining signal that can cause privacy leakage. The SILI values of all techniques along with their utility are shown in Figure 18. While most techniques (except PrivacyMic) show high utility, they have high SILI values showing high leakage. *FeatureSense* shows high utility along with low SILI score, thus preventing speaker attributes leakage while maintaining privacy-utility trade-off.

*8.3.2 Speech Privacy Leakage.* To evaluate speech leakage, we analyze the Mutual Information (MI) between the first three formants and various extracted audio features. According to the source-filter model of speech production, the first three formants (F1, F2, F3) are essential for capturing vowel information, which is critical for understanding speech content as discussed in Section 6. If extracted features exhibit high MI with these formants, it would indicate that they contain speech-related information, posing a privacy risk. Figure 19 shows that MI values remain low across all features, with only a few (such as Spectral Spread, Tonality Index, and Spectral Contrast) showing slightly higher values, though still below 0.22. This indicates that no single feature strongly correlates with all 3 formant frequencies, meaning the extracted features do not encode significant vowel or speech-related content. Since none of our extracted features show high MI with the first three formants, we conclude that the features used in our analysis do not leak speech information. This suggests that the current feature set can be employed in privacy-preserving audio analysis without compromising the intelligibility of spoken content.

## 8.4 Computational Cost Evaluation

The latencies for each feature were calculated by measuring the time taken to compute the feature during each sliding window iteration of the audio signal. These experiments are performed on Macbook M1 Pro. After processing all windows (each of size 500 ms), the average time per window for each feature was calculated by dividing the total time by the number of windows. For feature groups, the latencies were aggregated by averaging the average times of all features within the group.

**Optimizing Latency**: The extraction times reduced significantly after optimization due to the replacement of the `librosa` library with custom implementations using `numpy` and `scipy`. While `librosa` provides a high-level, user-friendly interface for audio processing, it often incurs additional computational overhead due to its generalized and feature-rich design. For instance, the cost of extracting RMS value using librosa was 1.2 ms per window, while it reduced to 0.031 ms per window by using numpy (97.4% reduction). The optimized implementation avoided these overheads by using lower-level, more direct mathematical operations tailored specifically to the required computations. This allowed for leaner and faster processing of the audio data, resulting in reduced latencies as shown in Figure 20.

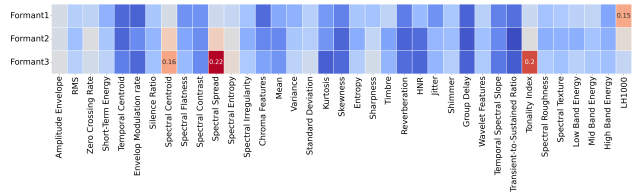

Fig. 19. Intelligibility Analysis of FeatureSense

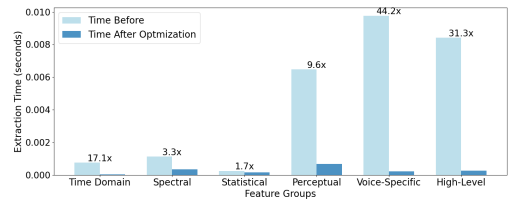

Fig. 20. Computational Latency of Feature Extraction

For example, FFT-based spectral computations and time-domain calculations were directly implemented using `numpy.fft` and mathematical formulas. Many operations that were previously loop-based (e.g., mapping frequencies to chroma bins) were replaced with vectorized operations using numpy. This significantly reduced the overhead of Python loops. Feature groups that relied heavily on librosa functions, such as Perceptual Features and Voice-Specific Features, saw the largest reductions in time. This is because librosa often performed additional, unnecessary computations that were eliminated in the optimized code. Groups like Spectral Features and Derived Features, which were simplified and vectorized using numpy, also showed significant reductions, as the custom implementations avoided redundant operations. Lastly, we use approximations for some features like reverberation was approximated using the decay of the envelope extracted via the Hilbert transform, which is computationally efficient. Similarly, HNR was computed as the ratio of the energy of the analytic signal (via Hilbert transform) to the noise energy. The approximations reduced the computational cost significantly by 81.4%.

**Raspberry Pi Evaluation**: To further analyze the cost on low end devices, we use Raspberry Pi 5 with a USB microphone. We perform the feature extraction time profiling and observed that while the total feature extraction latency on Laptop was 9.02 ms, it increased to 38.56 ms on Raspberry Pi (4.27x slower). Figure 21 shows the increase in feature extraction time from Macbook M1 Pro to Raspberry Pi 5.

Specifically, features like Spectral Texture observe a significant increase in the extraction time because it is using wavelet decomposition, which is computationally expensive raspberry Pi's lower computational power and memory bandwidth compared to a laptop make it slower. Similarly, reverbration calculation latency increased by 3.5x due to lack of advanced instruction sets in Pi. However, the overall feature extraction latency of all the features is **still less than 40 ms for a window size of 500 ms**, making is feasible for real-time applications.

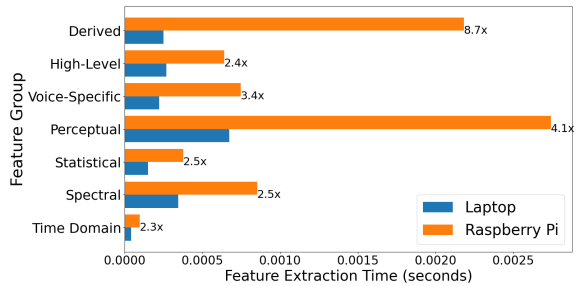

Fig. 21. Latency Profiling on Raspberry Pi

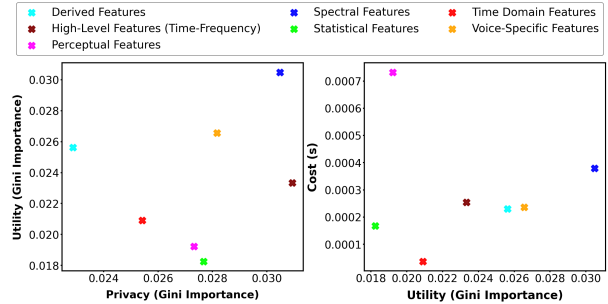

Fig. 22. Privacy-utility-cost Trade-offs for Different Feature Groups

This latency number can be further improved by parallel-processing across multiple CPU cores available on the Raspberry Pi, which can reduce the latency by 50%-75% depending on workload and number of CPU cores.

## 8.5 Privacy-Utility-Cost Trade-offs across Feature Groups

We observe that there is a clear trade-off between privacy, utility, and cost among different feature groups as shown in Figure 22. Features that offer high utility, such as Chroma Features, Envelope Modulation Rate, and Tonality Index, tend to have relatively higher privacy leakage. On the other hand, features with lower privacy leakage, such as Entropy, Silence Ratio, and Amplitude Envelope, often provide lower utility, indicating that strict privacy preservation may come at the cost of reduced effectiveness in tasks like classification or regression. When analyzing the Privacy vs. Utility plot (Figure 22), we see that some feature groups, such as Spectral Features and Voice-Specific Features, exhibit a high privacy leakage while maintaining strong utility, suggesting that they capture speaker-dependent characteristics. Conversely, Statistical Features and Time-Domain Features are more privacy-preserving but do not contribute as significantly to performance. We observe that the proposed derived features show high utility while leaking the least privacy as they were designed to capture various environment sound characteristics. The Utility vs. Cost plot reveals another crucial insight: high-utility features are often computationally expensive. For instance, HNR, and Spectral Texture require significant processing power, which may not be ideal for real-time applications. In contrast, Statistical Features like Mean, Variance, and Standard Deviation offer a reasonable trade-off between cost and utility, making them more suitable for low-power or embedded systems. From these insights, we can conclude that selecting an optimal feature set depends on the specific application. If maximizing model performance is the primary goal, we might prioritize high-utility features despite their privacy risks. However, if privacy preservation is crucial, we may need to sacrifice some performance by using lower-leakage features. Additionally, cost considerations suggest that certain high-utility but expensive features should only be used when computational resources allow. These results help us make informed decisions about feature selection based on the desired balance between privacy, utility, and cost.

## 8.6 Case Study 1: Privacy-Aware Cough Detection for Respiratory Health Apps.

Respiratory health applications such as Hyfe AI [41] use smartphone microphones to detect cough patterns and track respiratory conditions. However, many of these third-party apps currently receive access to raw audio, which allows not only cough detection but also unintended speaker profiling. For instance, an app could infer user attributes like ethnicity or gender and use that information for biased targeting, differential treatment, or advertising. Similarly, ambient health monitoring systems like AeroSense [3] aim to detect activities like coughing, sneezing, and speaking for estimating aerosol emissions in shared indoor spaces, yet these applications should not infer speaker identity or demographics. To support such privacy-respecting use cases, we propose integrating FeatureSense at the mobile OS or device driver level to expose only privacy-preserving features—ensuring accurate sound classification without leaking sensitive speaker traits. To evaluate this scenario, we train a cough-vs-non-cough binary classifier using labeled samples from the Google AudioSet dataset with a subset of Common Voice dataset. We then apply our feature selection algorithm with varying privacy-utility tradeoff parameter $\alpha \in \{0.3, 0.5, 0.8, 1.0\}$. As with the smart speaker use case, we fix a latency budget of $T < 0.15$ seconds to reflect real-time constraints for cough sensing and we select the task as "human non-speech sounds". The optimization algorithm identifies feature subsets that balance utility, privacy leakage (age attribute), and computation cost, given the task category. As shown in Table 7, utility retained decreases modestly as privacy increases. The raw audio model achieves an accuracy of 86.13% and an ethnicity classification accuracy of 66.1% (where random guessing among three classes yields ~33%). Using FeatureSense with $\alpha = 1.0$, the classification accuracy is 85.25%, which corresponds to a utility retention of 98.96% (i.e., 85.25 / 86.13). The corresponding ethnicity leakage drops to 38.9%, resulting in a 27.2% reduction compared to the raw audio baseline (i.e., 66.1 - 38.9). At $\alpha = 0.3$, utility retained remains relatively high at 92.17%, while ethnicity leakage is reduced by 41.6% compared to the raw audio baseline. This demonstrates that FeatureSense effectively balances privacy and utility—substantially lowering speaker attribute leakage while maintaining real-time performance and high task accuracy. Feature extraction latency drops significantly across the range—from 8.96 ms at $\alpha = 1.0$ to just 1.45 ms at $\alpha = 0.3$. Prediction latency remains stable around 9–10 ms, keeping the total latency under 15 ms across all privacy settings. These results demonstrate that FeatureSense supports real-time, privacy-preserving cough detection suitable for deployment in mobile and wearable health applications—offering a high-utility, low-latency solution while minimizing exposure of sensitive speaker demographic attributes such as ethnicity.

Table 7. Cough Detection on AudioSet (Raw Accuracy = 86.13%, Raw Ethnicity Leakage = 66.1%)

| $\alpha$ | # Features | Utility Retained (%) | Ethncity Leakage Reduction (%) | Feat. Lat. (ms) |
|---|---|---|---|---|
| 0.3 | 6 | 92.17 | 41.65 | 1.45 |
| 0.5 | 22 | 95.78 | 32.80 | 3.98 |
| 0.8 | 31 | 98.32 | 30.42 | 8.62 |
| 1.0 | 35 | 98.96 | 27.22 | 8.96 |

Table 8. UrbanSound8k Classification (Raw Accuracy = 97.64%, Raw Age Leakage = 63.8%)

| $\alpha$ | # Features | Utility Retained (%) | Age Leakage Reduction (%) | Feat. Lat. (ms) |
|---|---|---|---|---|
| 0.3 | 9 | 93.07 | 51.40 | 1.48 |
| 0.5 | 20 | 96.22 | 37.06 | 3.50 |
| 0.8 | 32 | 98.49 | 34.80 | 8.30 |
| 1.0 | 35 | 99.41 | 31.12 | 8.90 |

### 8.7 Case Study 2: Target Sound Detection with Smart Speaker Privacy Mode.

Imagine a scenario where a user enables "privacy mode" on a smart speaker (e.g., Alexa, Google Nest) before leaving home. In this mode, the device continues monitoring for critical environmental sounds—such as a dog barking, a gunshot, or an air conditioner anomaly—without capturing identifiable speaker traits like age, gender, or speech content. This prevents third-party services or cloud-based analytics from inferring whether a child or elderly individual is home alone, thus reducing risks of targeted advertising, surveillance, or manipulation. FeatureSense enables robust non-speech audio classification while explicitly minimizing speaker leakage in such settings. To evaluate this use case, we selected three relevant classes from the UrbanSound8k dataset: *dog bark*, *gun shot*, and *air conditioner*, mixed with a small subset of Common Voice dataset. We applied our privacy-aware feature selection pipeline while varying the privacy-utility tradeoff parameter $\alpha$ from 1.0 (maximum utility) to 0.3 (maximum privacy), under a latency constraint of $T < 0.15$ seconds. As shown in Table 8, lowering $\alpha$ reduces the number of features selected, resulting in increased privacy and lower computational cost, but modest degradation in classification accuracy. We report two key metrics: *Utility Retained*, computed as the classification accuracy relative to a accuracy of full audio spectrogram (97.64%), and *Age Leakage Reduction*, calculated as the percentage decrease in 3-class age classification accuracy compared to a raw audio baseline (age leakage = 63.8%). At $\alpha = 1.0$, FeatureSense retains 99.41% utility while reducing age leakage by 31.12%. At $\alpha = 0.3$, utility is still 93.07%, while age leakage drops by over 51.4%. Feature extraction latency drops significantly from 8.9 ms to 1.48 ms across this range, while prediction latency remains stable around 8–9 ms, keeping total latency under 15 ms across all settings. For comparison, the raw audio-based model achieves slightly higher classification accuracy (97.64%), but leaks substantially more speaker age information. This case study illustrates how FeatureSense enables configurable, low-latency, privacy-preserving audio sensing suitable for real-time smart home deployments.

## 9 DISCUSSION

In this section, we discuss additional aspects of audio privacy and some limitations of our work.

**Contextual Privacy**: As privacy depends on the context, it is non-trivial to measure privacy leakage. While this work takes an important step towards preserving the speaker's privacy, there remains the potential for unintended leakage of contextual information. For instance, background sounds such as traffic noise, birds chirping, or specific indoor ambiance (e.g., a cafe or office) could inadvertently reveal the speaker's location or surroundings. Additionally, temporal or sequential patterns in the audio data might disclose contextual clues about the speaker's activities or environment. These challenges highlight the inherent complexity of achieving comprehensive privacy, as context-related information is deeply intertwined with audio signals. It is therefore inevitable that achieving absolute privacy remains a nuanced and ongoing challenge, requiring further research. This work is an important step in this direction as we show that for a given task, you can choose the attribute that you want to preserve.

**Feature List Expansion**: Although audio features is a non-exhaustive list, we cover most of the possible feature categories and pick the ones that are privacy preserving. We understand that it is still possible to design more derived audio features that might be effective. We provide an easy-to-use expansible library where we can add more features. In this work, we explore a comprehensive list of features. The framework is modular, and we plan to release a training interface for people to derive their own privacy-aware feature sets and evalue it in terms of privacy, utility, and cost. Future work can explore automatic feature generation [53] using attention maps or LLM-based approaches. Then, we can assess the privacy of those features using the proposed approach and penalize the selection of privacy-invasive features.

**Assessing Speech Leakage of Granular Features**: Evaluating the speech leakage of granular and localized features is challenging. Since we do not have access to the audio signal or spectrogram, we cannot perform human intelligibility tests or ASR-based tests to get the WER. In this work, we use correlation analysis with formants for understanding speech leakage. For future work, we plan to address this challenge by using GAN-based speech synthesizers that would regenerate the candidate audio signal(s) and perform speech leakage evaluation on it.

**Extensive Utility Assessment**: We show high utility across various tasks like environments sounds classification, urban sounds detection, and cough detection. However, we plan to further evaluate the proposed system in various application domains to understand the potential of this system. For example, speech/non-speech detection, speaker change detection, voice pathology detection. We understand that the proposed approach might not work for tasks like speech recognition, speaker recognition, and emotion detection, due to its privacy-first design approach.

**Residual Speaker Leakage and Adversarial Risks:** We acknowledge that while FeatureSense significantly reduces speaker identity leakage, it may not entirely eliminate all identifiable traits, leaving room for sophisticated adversarial de-anonymization attacks. To enhance resilience against such threats, we are exploring the integration of adversarial training techniques that dynamically identify and suppress residual identity-bearing features. Additionally, incorporating differential privacy mechanisms into our feature extraction process can provide formal guarantees on the upper bounds of information leakage, further strengthening privacy protections.

**Adaptive Feature Granularity:** Currently, FeatureSense operates on fixed-length 500 ms audio windows for feature extraction and selection. However, some applications—such as detecting persistent background sounds or periodic environmental events—may tolerate larger window sizes without loss of performance. Incorporating window duration as an optimization parameter in the privacy-utility-cost tradeoff framework can further restrict the temporal granularity of information shared with third-party services. This not only reduces the resolution of potentially sensitive acoustic cues but also enhances privacy by minimizing the amount of audio context exposed per inference.

**Audio Device Driver Layer Deployment:** We envision FeatureSense as a modular privacy layer deployed within the audio device driver stack, enabling privacy-preserving decisions to be made locally and in real-time. At this layer, the system could assess the privacy sensitivity of features dynamically and suppress speaker-revealing attributes before audio data is processed by downstream applications. Furthermore, this approach allows context-aware control—such as activating stricter privacy filtering when the user is in sensitive environments—without requiring application-level intervention. This positions FeatureSense as a practical and scalable solution for always-on sensing on smartphones, smart speakers, and embedded IoT systems.

## 10 RELATED WORK

Ensuring privacy in audio sensing has been an important research problem. In the United States, privacy laws are evolving to address concerns related to voice data. A comprehensive review by Dutta and Hansen (2024) examines existing and proposed legislation, highlighting considerations for processing children's data and the role of synthetic data in AI applications [54]. Existing works try to degrade the audio signal in multiple ways to preserve privacy-sensitive content like sub-sampling, filtering, and injecting noise. There has been a rich literature in suppressing speech to preserve privacy. This is typically done by detecting likely speech or voiced segments using a Voice Activity Detection (VAD) [55] or designing a classifier [1, 11]. These detected segments are them eliminated [18, 56], obfuscated with noise [53], or replaced with pre-recorded signal [33].

**Preventing Speaker Attributes Leakage**: Recent research motivates the need to look at privacy beyond speech [12]. Voice data can be used not only for purposes such as speech characterization, but also for applications that characterize the speaker [57]. Some works have considered privacy apart from speech. For example, Testa et al. [9] added noise to mask users emotional information while preserving the transcription-relevant portions of their speech. Emotionless [58] prevented sensitive emotional state leakage of the speaker is by 96 %. Additionally, there have been efforts to protect speaker biometric identity [59]. These works usually measure speaker identity leakage using EER metrics. However, there is limited work in preserving and assessing speaker's demographic attributes leakage that can significantly cause privacy harm. Alsenani et al. [?] conducted a systematic study showing that speech emotion recognition systems can leak gender information with inference accuracies as high as 95%. Beyond audio, Shi et al. [60] demonstrated that facial motion data captured by AR/VR headsets can be exploited to infer speaker identity and demographics without microphone access. These works underscore the urgency of mitigating speaker attribute leakage.

**Different levels of privacy**: Existing audio privacy techniques span hardware-level, signal-level, feature-level, and model-level interventions. Table 9 summarizes these approaches, highlighting their challenges and privacy evaluation methodologies. Several recent works propose hardware- and model-layer approaches to protect audio privacy. MicPro [61] introduces a hardware design that physically filters the microphone signal to suppress sensitive voice content. Learning Normality is Enough [62] proposes a model-layer defense against inaudible voice attacks by learning typical audio patterns. SafeEar [63] focuses on detecting audio deepfakes through content-based privacy protection. In contrast, FeatureSense offers a software-based, lightweight alternative that operates directly at the feature extraction stage, enabling real-time privacy-preserving sensing without requiring new hardware or complex deep learning models. This modularity makes it practical for integration into a wide range of sensing pipelines without needing changes to hardware or ML models.

Table 9. Audio Privacy Techniques Across System Levels

| Level | Technique | Challenges | Privacy Evaluation Methods |
| --- | --- | --- | --- |
| Hardware-level | Physical signal perturbation or filtering [61] | Requires hardware modification; limits flexibility | ASR methods (WER, PER) or human intelligibility study |
| Signal-level | Noise addition or obfuscation [56] | Speech still partially leaks; degrades utility | ASR methods (WER, PER) or human intelligibility study |
| Signal-level | Selective sensing [11] | Relies on VAD accuracy; speech segments may still pass | ASR methods (WER, PER) or human intelligibility study |
| Signal-level | Filtering or subsampling [16, 29] | Tradeoff between utility and privacy depends on cut-off frequency | ASR methods (WER, PER) or human intelligibility study |
| Feature-level | Adversarial feature learning [64, 65] | Complex training; sensitive to discriminator strength | Speaker identification attacks (EER) |
| Feature-level | Low-level granular feature selection [24, 35, 36, 66, 67] | Difficult to formally evaluate leakage; non-invertibility helps | Hard to evaluate, Proposed metric: *SILI* |
| Model-level | Differential Privacy (DP), Federated Learning (FL) [68] | Noise degrades model performance significantly, Reduces raw data sharing but gradients can leak info | Privacy budget analysis (epsilon), FL-specific inversion attacks |

**Feature-Based approaches**: Sankaran et al. [69] provides a comprehensive survey of audio features (including time-domain, frequency-domain, and perceptual) that have been used for the detection of voice pathology and showed good results with MFCC features. Similarly, [4] compares different feature representations (like MFCC, Mel spectrogram, STFT, etc.) for cough detection and found that MFCC shows the best accuracy. Even though MFCCs show good accuracy on wide-range of applications, they are known to leak speaker-related privacy [49] and have been used in the literature for speaker recognition tasks [70]. This enhances the need for exploring features that maintain the effectiveness of audio classification tasks while addressing privacy concerns. Feature-based techniques have become increasingly popular in audio sensing applications due to their efficiency, low latency, and competitive accuracy [24, 35, 36, 66]. However, **most of these systems do not explicitly assess the privacy risks associated with their chosen feature sets**, particularly in terms of speaker demographic or speech content leakage. Moreover, **designing effective and privacy-conscious features often requires significant domain expertise**, which limits the accessibility and generalizability of such approaches. *FeatureSense* **addresses both of these limitations by systematically evaluating feature-level leakage and providing an adaptive framework that balances privacy, utility, and computational cost**—enabling developers to build privacy-aware audio systems without requiring deep acoustic knowledge.

## 11 CONCLUSION

As microphones are ubiquitous, it is crucial to ensure trust before deploying audio-based applications on smartphones and smart speakers. We propose a privacy evaluation metric that anyone can use to assess the privacy of their system. We propose *FeatureSense* library that exposes a list of privacy-aware features that increase the accuracy of audio classification without requiring domain knowledge. We show that the proposed library extracts these features in real-time. We present an optimization algorithm that penalizes the selection of privacy-invasive features for extremely sensitive application. The resulting adaptive feature list can be used for enabling context-aware applications. The proposed approach prevents speaker-related privacy leakage by 60.6% while maintaining the effectiveness of audio-based applications. By designing a comprehensive privacy evaluation framework and adaptive task-specific feature selection, this work provides a foundational framework for ensuring trust in audio sensing technologies. Through this work, we aim to push the boundaries of audio sensing, setting the foundations for future innovation in the field.

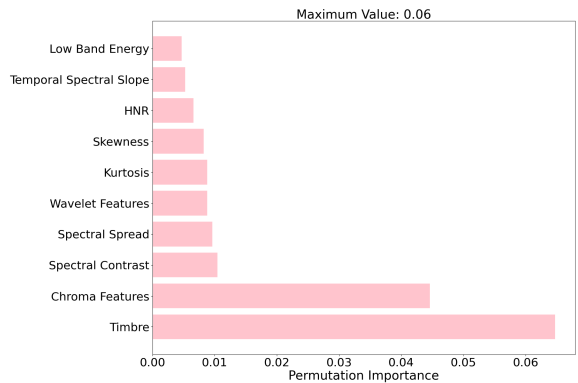

Fig. 23. Top 10 Individual Permutation Importance

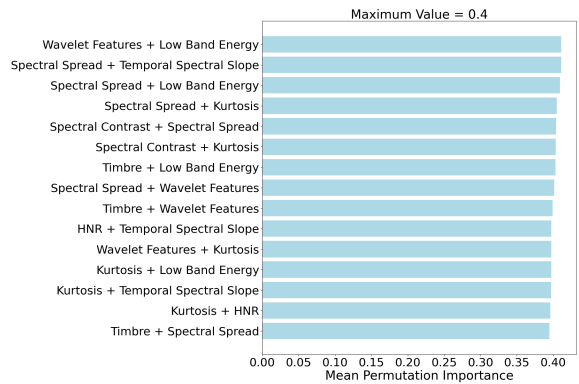

Fig. 24. Top 10 Combined Permutation Importance of Feature Pairs

## A APPENDIX

**Privacy Leakage of Individual Features and their Combinations** Some features may appear uncorrelated individually but, when combined, could predict sensitive attributes. The interaction between multiple features might contribute to the leakage, which correlation and MI analysis doesn't capture. So, we evaluated PI to assess the predictive contribution of individual features versus their combinations to the sensitive attribute like gender. For individual features, importance values were computed by randomly shuffling feature values and measuring the drop in model performance. Similarly, combinations were analyzed by considering pairs of features to capture interaction effects. Figure 23 shows the top 10 features with highest permutation importance score. Next, we form all possible pairs of these 10 features and extract their mean permutation importance. Figure 24 shows the 10 best performing combinations. The results indicate that while individual features, such as Timbre and Chroma Features and Spectral Contrast, showed relatively high predictive power, feature combinations (e.g., Low Band Enegy + Wavelet Features) exhibited significantly higher permutation importance (6.6x). This suggests that individual features may not leak much information alone, but their interactions can uncover patterns that contribute to leakage of sensitive speaker attribute, emphasizing the importance of analyzing feature combinations when assessing privacy risks. Another observation here is that although Timbre and Chroma features had high individual PI score, but their combination was not the best performing PI combination. So, the features that individually do not leak significant privacy, their combinations could still lead to high leakage. So, we train Random Forest Classification models for age, gender, ethnicity with all the features and get feature importance rankings. We provide these rankings available through *FeatureSense* library as described in Sectio 7.